# A Possible Theoretical Basis for Propulsive Force Generation by Both Conventional and Unconventional Means

C. C. Briggs
*Center for Academic Computing, Penn State University, University Park, PA 16802*

**Abstract.** A possible theoretical basis is given for propulsive force generation by both conventional and unconventional means.

PACS Numbers: 04.50.+h

This paper presents a possible theoretical basis for propulsive force generation by both conventional and unconventional means.

The theoretical basis for conventional means of propulsive force generation (e.g., rocket engines) is contained—partly because of allowances having been made for possible variations of various mathematical quantities able to encompass various potentially controllable physical quantities (e.g., exhaust velocity) having non-zero functional dependences on spacetime or other coordinates or parameters—in "equations of motion"[1-4] derivable from 4-dimensional energy-momentum conservation equations, which are themselves derivable in general relativity from the covariant divergence of a 4-dimensional energy-momentum (or "matter") tensor, which, by Einstein's equations, is parallel to the 4-dimensional Einstein curvature tensor, which satisfies a 2*nd* contraction of Bianchi's identities for the Riemann-Christoffel curvature tensor, various terms in the equations being interpretable as thrust.

The generalized energy-momentum conservation equations derived from the covariant divergence of an *n*-dimensional energy-momentum tensor are similar to the aforementioned 4-dimensional conservation equations and also contain terms interpretable as thrust. Extra, "correction" terms appear in the equations as a result of (1) the use of geometries of a higher dimensionality than the usual four (e.g., five[5]), (2) the use of less restrictive geometries than the usual Riemannian (e.g., geometries employing non-torsion-free[6] or non-metrical[7] connections), and (3) the accommodation of extra functional dependences (e.g., dependences on a 5*th* dimension[8] supplementary to space and time). Those extra terms may (1) indicate new means of generating propulsive forces as well as (2) serve to explain a variety of other, evidently still hypothetical unconventional means of generating propulsive forces directly (e.g., electromagnetically or otherwise, but without using propellants[9-37]) or indirectly (e.g., by altering inertial masses electromagnetically[38-44]).

---

[1] Ackeret, J., "Zur Theorie der Raketen," *Helvetica Physica Acta*, **19** (1946) 103.
[2] Bade, W. L., "Relativistic Rocket Theory," *Amer. J. Phys.*, **21** (1953) 310.
[3] Pomeranz, K. B., "Newton's Law for Systems with a Variable Mass," *Amer. J. Phys.*, **32** (1964) 386; "The Equations of Motion for Relativistic Particles and Systems with a Variable Rest Mass," *Amer. J. Phys.*, **32** (1964) 955; "The Relativistic Rocket," *Amer. J. Phys.*, **34** (1966) 565; "Some Relativistic Effects in Systems with a Variable Rest Mass," *Amer. J. Phys.*, **37** (1969) 741, "Erratum: *Some Relativistic Effects in Systems with a Variable Rest Mass*," **38** (1970) 119.
[4] Rhee, J. W., "Relativistic Rocket Motion," *Amer. J. Phys.*, **33** (1965) 587.
[5] Williams, P. E., *The Dynamic Theory: A New View of Space, Time, and Matter*, Los Alamos Scientific Laboratory, Department of Defense, Washington, DC (1980), National Technical Information Service report no. LA-8370-MS; *The Dynamic Theory: Some Shock Wave and Energy Implications*, Los Alamos Scientific Laboratory, Department of Defense, Washington, DC (1981), National Technical Information Service report no. LA-8402-MS; *The Arrow of Time in the Dynamic Theory*, Los Alamos Scientific Laboratory, Department of Defense, Washington, DC (1981), National Technical Information Service report no. LA-8690-MS; *The Possible Unifying Effect of the Dynamic Theory*, Los Alamos National Laboratory, Department of Defense, Washington, DC (1983), National Technical Information Service report no. LA-9623-MS.
[6] Bressan, O. J., "Classical Unified Fields: Physical Space," *Gen. Rel. Grav.*, **18** (1986) 375.
[7] Pullin, J., and O. Bressan, "New Weyl Theory: Geometrization of Electromagnetism and Gravitation: Motivations and Classical Results," *Gen. Rel. Grav.*, **19** (1987) 601.
[8] Williams, P. E., *op. cit.*
[9] Holt, A. C., *Field resonance propulsion concept*, National Aeronautics and Space Administration, Washington, DC (1979), NASA Technical Reports report no. NASA-TM-80961, JSC-16073, document ID no. 19800010907 N (80N19184); "Prospects for a breakthrough in field-dependent 'propulsion,'" in *Proceedings of the 16th AIAA/SAE/ASME Joint Propulsion Conference, June 30-July 2, 1980/Hartford, CT*, American Institute of Aeronautics and Astronautics, Washington, DC (1980), paper no. AIAA-80-1233.
[10] [Author(s) Not Available], *Gravitational propulsion*, National Aeronautics and Space Administration, Washington, DC (1988), NASA Technical Reports document ID no. 19880019412 N (88N28796).
[11] Guay, B. T., "Propulsion Without Propellant Mass; a Time-Varying Electromagnetic Field Effect," Los Alamos preprint no. physics/9908048; "About Charge Density Wave [sic] for Electromagnetic Field-Drive," Los Alamos preprint no. physics/9911061.
[12] Vance, J. H., "A method of deep space propulsion," *Journal of Practical Applications in Space*, **2** (1991) 105.
[13] Alcubierre, M., "The warp drive: hyper-fast travel within general relativity," *Class. Quantum Grav.*, **11** (1994) L73.
[14] Haloulakos, V. E., "Exotic propulsion systems—A space exploration imperative," in *Proceedings of the 28th AIAA/ASME/SAE/ASEE Joint Propulsion Conference and Exhibit, July 6-8, 1992/Nashville, TN*, American Institute of Aeronautics and Astronautics, Washington, DC (1992), paper no. AIAA-92-3779; "21st century propulsion—Key issues and some answers," in *Proceedings of the 29th AIAA/ASME/SAE/ASEE Joint Propulsion Conference & Exhibit, June 28-30, 1993/Monterey, CA*, American Institute of Aeronautics and Astronautics, Washington, DC (1993), paper no. AIAA-93-1810.
[15] Noyes, H. P., "Anti-gravity: The key to 21st century physics," National Aeronautics and Space Administration, Washington, DC (1993), NASA Technical Reports report no. DE93-006929, SU-SLAC-PUB-5857, CONF-9209297-1, document ID no. 19930015274 N (93N24463); cf. "Anti-Gravity," (1992), Stanford Linear Accelerator Center preprint no. SLAC-PUB-5990.
[16] Clark, C., W. A. Hiscock, and S. L. Larson, "Null geodesics in the Alcubierre warp drive spacetime: the view from the bridge," Montana State University preprint no. MSUPHY99.02, Los Alamos preprint no. gr-qc/9907019.
[17] Cassenti, B. N., and H. I. Ringermacher, "The 'How To' of Antigravity," in *Proceedings of the 32nd AIAA/ASME/SAE/ASEE Joint Propulsion Conference, July 1-3, 1996/Lake Buena Vista, FL*, American Institute of Aeronautics and Astronautics, Washington, DC (1996), paper no. AIAA-96-2788.
[18] Cravens, D. L., *Electric Propulsion Study*, AD-A227 121, prepared for the Astronautics Laboratory, Air Force Space Technology Center, Space Systems Division, Air Force Systems Command, Edwards Air Force Base, CA 93523-5000, Department of Defense, Washington, DC (1990), Defense Technical Information Center report no. AL-TR-89-040.
[19] Talley, R. L., *Twenty First Century Propulsion Concept* [sic], AD-A237 853, prepared for the Phillips Laboratory, Air Force Systems Command, Propulsion Directorate, Edwards Air Force Base, CA 93523-5000, Department of Defense, Washington, DC (1991), Defense Technical Information Center report no. PL-TR-91-3009.
[20] Van Den Broeck, C., "A 'warp drive' with more reasonable total energy requirements," Katholieke Universiteit Leuven preprint no. KUL-TF-99/18, Los Alamos preprint no. gr-qc/9905084; "On the (im)possibility of warp bubbles," (1999), Katholieke Universiteit Leuven preprint no. KUL-TF-99/22, Los Alamos preprint no. gr-qc/9906050.
[21] Forward, R. L., "Space Warps: A Review of One Form of Propulsionless Transport," *Journal of The British Interplanetary Society*, **42** (1989) 533; "Negative Matter Propulsion," *Journal of Propulsion and Power*, **6** (1990) 28.
[22] González-Díaz, P. F., "On the warp drive space-time," Los Alamos preprint no. gr-qc/9907026.
[23] Millis, M. G., "Exploring the Notion of Space Coupling Propulsion," in *Vision 21: Space Travel for the Next Millennium*, Symposium Proceedings, April 1990, NASA-CP-10059, National Aeronautics and Space Administration, Washington, DC (1990), p. 307; "Coupling Gravity, Electromagnetism and Space-Time for Space Propulsion Breakthroughs," in *Proceedings of the Sixth Annual Symposium of the NASA Propulsion Engineering Center of the Pennsylvania State University Held at NASA Lewis Research Center, Cleveland, Ohio, September 13-14, 1994*, National Aeronautics and Space Administration, Washington, DC (1994), p. 86; "The challenge to create the space drive," National Aeronautics and Space Administration, Washington, DC (1996), NASA Technical Reports report no. NASA-TM-107289, NAS 1.15:107289, E-10365, document ID no. 19960048672 N (96N33893); "Challenge to Create the Space Drive," *Journal of Propulsion and Power*, **13** (1997) 577; "Breakthrough Propulsion Physics Research Program," National Aeronautics and Space Administration, Washington, DC (1996), [prepared for the Space Technology and Applications International Forum *STAIF-97: Looking Forward: Opening the Space Frontier* cosponsored by NASA, DSWA, U.S. DOE, and USAF, Albuquerque, New Mexico, January 26-30, 1997], NASA Technical Memorandum no. 107381, NASA Technical Reports report no. NASA-TM-107381, E-10569, NAS 1.15:107381.
[24] Froning, H. D., Jr., and R. L. Roach, "Field propulsion for future needs," in *Proceedings of the 28th AIAA/ASME/SAE/ASEE Joint Propulsion Conference and Exhibit, July 6-8, 1992/Nashville, TN*, American Institute of Aeronautics and Astronautics, Washington, DC (1992), paper no. AIAA-92-3780; "Drag Reduction, and Possibly Impulsion, by Perturbing Fluid and Vacuum Fields," in *Proceedings of the 31st AIAA/ASME/SAE/ASEE Joint Propulsion Conference and Exhibit, July 10-12, 1995/San Diego, CA*, American Institute of Aeronautics and Astronautics, Washington, DC (1995), paper no. AIAA-95-2368.
[25] Bennett, G. L., and R. H. Frisbee, "Summary of the NASA/JPL Workshop on Advanced Quantum/Relativity Theory Propulsion," in *AIP Conference Proceedings of the Space Technology and Applications International Forum (STAIF-97), January 26-30, 1997, Albuquerque, NM* [*STAIF-97: Looking Forward: Opening the Space Frontier*], American Institute of Physics, Woodbury, NY (1997), p. 1515.
[26] Bennett, G. L., R. L. Forward, and R. H. Frisbee, "Summary of the NASA/JPL Workshop on Advanced Quantum/Relativity Theory Propulsion," in *Proceedings of the 31st*

Expressions for 5-dimensional connection coefficients and corresponding curvature tensors accommodating arbitrary functional dependences on the $5^{th}$ dimension (using holonomic coordinates for a non-torsion-free, metrical linear connection) are given in Appendix A (see below).

Covariant generalizations of the usual geodesic and geodesic deviation (or "tidal force") equations represent two kinds of "force equations," viz., (1) equations defining a "ponderomotive force" density vector ${F_{(1)}}^a$ derivable from the conservation equations, i.e.,

$$ {F_{(1)}}^a \equiv \nabla_b T^{ab}, \tag{1} $$

from which generalized geodesic equations for trajectories can be derived, and (2) equations defining a "tidal force" density vector ${F_{(2)}}^a$ derivable from the geodesic-deviation equations for trajectories, i.e.,

$$ {F_{(2)}}^a \equiv \mu\, u^b\, \nabla_b u^c\, \nabla_c v^a, \tag{2} $$

where the scalar $\mu$ represents mass density, the vector $u^a$ the tangent vector of the trajectory (a "generalized velocity vector"), and the vector $v^a$ the deviation vector of the trajectory.

Generalized geodesic equations (which, in typical versions of 5-dimensional general relativity, automatically include the Lorentz or "Lorentz-Heaviside" electromagnetic ponderomotive force equations) can be derived from the generalized energy-momentum (and, in five or more dimensions, electric charge-current, etc.) conservation equations as expressed in terms of—and usually by the vanishing of—the covariant divergence $\nabla_b T^{ab}$ of the energy-momentum tensor $T^{ab}$, where

$$ \nabla_b T^{ab} = \partial_b T^{ab} + {\Gamma_b}^a{}_c T^{cb} + {\Gamma_b}^b{}_c T^{ac}, \tag{3} $$

where ${\Gamma_a}^b{}_c$ is the connection coefficient and $\partial_a$ is the Pfaffian derivative.

The usual 4-dimensional Einstein gravitational field equations are given by

$$ G^{ab} = R^{ab} - \tfrac{1}{2} g^{ab} R = k\, T^{ab}, \tag{4} $$

where $G^{ab}$ is the Einstein curvature tensor, $R^{ab}$ is the Ricci curvature tensor, $R$ is the Riemann curvature scalar, and $k$ is a constant, which may be taken to be given by

$$ k = \frac{8\pi G}{c^2}, \tag{5} $$

where $G$ is Newton's gravitational constant and $c$ is the speed of light in vacuum.

The contravariant Einstein curvature tensor $G^{ab}$ satisfies the identity

$$ \begin{aligned} \nabla_b G^{ab} = {} & {Q_b}^{ac} {G_c}^b + (\tfrac{1}{2} {Q_b}^c{}_d - {S_{bd}}^c)\, {R^a}_c{}^{db} + {Q^{[b}}_{bd} R^{a]d} - \\ & - 2 {S_b}^{ac} {R_c}^b + g^{ab} \nabla_d (\nabla_{[c} {Q_{b]}}^{cd} + {S_{cb}}^e {Q_e}^{cd}) + \\ & + 2 {S_b}^{ac} (\nabla_{[d} {Q_{c]}}^{db} + {S_{dc}}^e {Q_e}^{db}), \end{aligned} \tag{6} $$

where ${Q_a}^{bc}$ is the non-metricity tensor and ${S_{ab}}^c$ the torsion tensor, as shown in Appendix B (see below).

Thus, using Eqs. (1), (3), (4), and (6), ${F_{(1)}}^a$ is given by

$$ \begin{aligned} {F_{(1)}}^a &= \nabla_b T^{ab} \\ &= \partial_b T^{ab} + {\Gamma_b}^a{}_c T^{cb} + {\Gamma_b}^b{}_c T^{ac} \\ &= \frac{1}{k} \nabla_b G^{ab} \end{aligned} \tag{7} $$

*AIAA/ASME/SAE/ASEE Joint Propulsion Conference and Exhibit, July 10-12, 1995/San Diego, CA*, American Institute of Aeronautics and Astronautics, Washington, DC (1995), paper no. AIAA-95-2599.

[27] Froning, H. D., Jr., "Experiments to Explore Space Coupling by Specially Conditioned Electromagnetic Fields," National Aeronautics and Space Administration, Washington, DC (1999), NASA Technical Reports document ID no. 19990023228, in *NASA Breakthrough Propulsion Physics Workshop Proceedings*, p. 207.

[28] Froning, H. D., Jr., and T. W. Barrett, "Inertia Reduction—and Possibly Impulsion—by Conditioning Electromagnetic Fields," in *Proceedings of the 33rd AIAA/ASME/SAE/ASEE Joint Propulsion Conference & Exhibit, July 6-9, 1997/Seattle, WA*, American Institute of Aeronautics and Astronautics, Washington, DC (1997), paper no. AIAA-97-3170.

[29] Graneau, P., "Electromagnetic jet-propulsion in the direction of current flow," *Nature*, **295** (1989) 311.

[30] Brandenburg, J. E., and J. F. Kline, "Application of the Gravity-Electro-Magnetism Unification Theory to the Problem of Controlled Gravity: Theory and Experiment," in *Proceedings of the 34th AIAA/ASME/SAE/ASEE Joint Propulsion Conference & Exhibit, July 13-15, 1998/Cleveland, OH*, American Institute of Aeronautics and Astronautics, Washington, DC (1998), paper no. AIAA-98-3137.

[31] Froning, H. D., Jr., T. Barrett, and G. Hathaway, "Preliminary Experiments Involving: Specially Conditioned EM Radiation, Matter, and Spacetime Metric," in *Proceedings of the 34th AIAA/ASME/SAE/ASEE Joint Propulsion Conference & Exhibit, July 13-15, 1998/Cleveland, OH*, American Institute of Aeronautics and Astronautics, Washington, DC (1998), paper no. AIAA-98-3138.

[32] Ringermacher, H. I., and B. N. Cassenti, "An Experiment to Verify a Relativistic Field Theory Based on an Electrodynamic Connection," in *Proceedings of the 34th AIAA/ASME/SAE/ASEE Joint Propulsion Conference & Exhibit, July 13-15, 1998/Cleveland, OH*, American Institute of Aeronautics and Astronautics, Washington, DC (1998), paper no. AIAA-98-3136.

[33] Petkov, V., "On the Possibility of a Propulsion Drive Creation Through a Local Manipulation of Spacetime Geometry," in *Proceedings of the 34th AIAA/ASME/SAE/ASEE Joint Propulsion Conference & Exhibit, July 13-15, 1998/Cleveland, OH*, American Institute of Aeronautics and Astronautics, Washington, DC (1998), paper no. AIAA-98-3142; "Propulsion Through Electromagnetic Self-Sustained Acceleration," *ibid.*, paper no. AIAA-99-2144; "Acceleration-Dependent Electromagnetic Self-Interaction Effects as a Basis for Inertia and Gravitation," Los Alamos preprint no. physics/9909019.

[34] Davis, E. W., "Wormhole Induction Propulsion (WHIP)," National Aeronautics and Space Administration, Washington, DC (1999), NASA Technical Reports document ID no. 19990023222, in *NASA Breakthrough Propulsion Physics Workshop Proceedings*, p. 157.

[35] Eagleton, R. D., M. N. Kaplan, and N. W. Kantor, "Force Field Propulsion," in National Aeronautics and Space Administration, Washington, DC (1999), NASA Technical Reports document ID no. 19990023232, in *NASA Breakthrough Propulsion Physics Workshop Proceedings*, p. 235.

[36] Forward, R. L., "Antigravity," *Proc. of the IRE*, **49** (1961) 1442; "Guidelines to Antigravity," *Amer. J. Physics*, **31** (1963) 166; see also "General Relativity for the Experimentalist," *Proc. of the IRE*, **49** (1961) 892.

[37] Fenioux, A., "Spatial motion without material propulsion," National Aeronautics and Space Administration, Washington, DC (1980), NASA Technical Reports report no. NASA-TM-75933, document ID no. 19800017885 N (80N26384).

[38] Matthews, R., "Inertia: Does Empty Space Put Up the Resistance?" *Science*, **263** (1994) 613.

[39] Puthoff, H. E., "Can the Vacuum Be Engineered for Spaceflight Applications? Overview of Theory and Experiments," *Journal of Scientific Exploration*, **12** (1998) 295.

[40] Haisch, B., A. Rueda, and H. E. Puthoff, "Inertia as a zero-point-field Lorentz force," *Phys. Rev. A*, **49** (1994) 678; "Beyond $E = m c^2$; A first glimpse of a postmodern physics, in which mass, inertia and gravity arise from underlying electromagnetic processes," *The Sciences*, **34** (1994) 26; "Physics of the Zero-Point Field: Implications for Inertia, Gravitation and Mass," *Speculations in Science and Technology*, **20** (1997) 99; "Advances in the Proposed Electromagnetic Zero-Point Field Theory of Inertia," in *Proceedings of the 34th AIAA/ASME/SAE/ASEE Joint Propulsion Conference & Exhibit, July 13-15, 1998/Cleveland, OH*, American Institute of Aeronautics and Astronautics, Washington, DC (1998), paper no. AIAA-98-3143.

[41] Haisch, B., and A. Rueda, "A Quantum Broom Sweeps Clean," *Mercury*, **25** (1996) 12; "The Zero-Point Field and the NASA Challenge to Create the Space Drive," in *Proceedings of the NASA Breakthrough Propulsion Physics Workshop, NASA Lewis Research Center, Aug. 12-14, 1997*; "Inertia as reaction of the vacuum to accelerated motion," *Physics Letters*, **240A** (1998) 115; "An Electromagnetic Basis for Inertia and Gravitation: What Are the Implications for 21st Century Physics and Technology?" in El-Genk, M. S. (ed.), *Space Technology and Applications International Forum-1998*, DOE CNF-980103, CP420 (1998), p. 1443; "The Zero-Point Field and Inertia," in Hunter, G., S. Jeffers, and J.-P. Vigier (eds.), *Causality and Locality in Modern Physics*, Kluwer Academic Publishers, Dordrecht, The Netherlands (1998), p. 171; "Progress in Establishing a Connection Between the Electromagnetic Zero-Point Field and Inertia," in *AIP Conference Proceedings of the Space Technology and Applications International Forum (STAIF-99), January 31-February 4, 1999, Albuquerque, NM*, American Institute of Physics, Woodbury, NY (1999), Los Alamos preprint no. gr-qc/9906069; "Toward an Interstellar Mission: Zeroing In on the Zero-Point-Field Inertia Resonance," in *AIP Conference Proceedings of the Space Technology and Applications International Forum (STAIF-2000), Conference on Enabling Technology and Required Scientific Developments for Interstellar Missions, January 30-February 3, 2000, Albuquerque, NM*, Los Alamos preprint no. physics/9909043.

[42] Forward, R. L., "Mass Modification Experiment Definition Study (An Air Force Report)," *Journal of Scientific Exploration*, **10** (1996) 325.

[43] Agop, M., C. G. Buzea, and B. Ciobanu, "On Gravitational Shielding in Electromagnetic Fields," Los Alamos preprint no. physics/9911011.

[44] De Aquino, F., "The Gravitational Spacecraft," *Electric Spacecraft Journal*, **27** (1998) 6, Los Alamos preprint no. physics/9904018; "The Correlation Between Gravitation and Electromagnetism, Inertia, and Unification," Los Alamos preprint no. physics/9905003; "Gravitation, Electromagnetism and Superparticles in the Initial Universe," Los Alamos preprint no. gr-qc/9905050; "Gravitation and Electromagnetism; Correlation and Grand Unification," Los Alamos preprint no. gr-qc/9910036.

$$= \frac{1}{k}\,\partial_b\, G^{ab} + \frac{1}{k}\,\Gamma_{b\ c}^{\ a}\, G^{cb} + \frac{1}{k}\,\Gamma_{b\ c}^{\ b}\, G^{ac}$$

$$= \frac{1}{k}\, Q_b^{\ ac}\, G_c^{\ b} + \frac{1}{k}\,(\tfrac{1}{2}\, Q_{b\ d}^{\ c} - S_{bd}^{\ \ c})\, R^{a\ db}_{\ c} + \frac{1}{k}\, Q^{[b}_{\ \ bd}\, R^{a]d} - \frac{2}{k}\, S_b^{\ ac}\, R_c^{\ b} +$$

$$+ \frac{1}{k}\, g^{ab}\, \nabla_d\, (\nabla_{[c}\, Q_{b]}^{\ \ cd} + S_{cb}^{\ \ e}\, Q_e^{\ cd}) + \frac{2}{k}\, S_b^{\ ac}\, (\nabla_{[d}\, Q_{c]}^{\ \ db} + S_{dc}^{\ \ e}\, Q_e^{\ db}).$$

Generalizations of the usual geodesic deviation equations are given (cf. Hayden[45]) by

$$u^b\ \nabla_b\, u^c\ \nabla_c\, v^a = u^b\, v^c\, u^d\, (R_{bcd}^{\ \ \ a} - 2\ \nabla_b\, S_{cd}^{\ \ a}) + 2\, u^b\, u^c\, S_{bd}^{\ \ a}\ \nabla_c\, v^d \tag{8}$$

$$= u^b\, v^c\, u^d\, (R_{bcd}^{\ \ \ a} - 2\ \nabla_b\, S_{cd}^{\ \ a} - 4\, S_{be}^{\ \ a}\, S_{cd}^{\ \ e}) +$$

$$+ 2\, u^b\, v^c\, S_{bd}^{\ \ a}\ \nabla_c\, u^d$$

as shown in Appendix C (see below).

Thus, using Eqs. (2) and (8), the tidal force density vector $F_{(2)}^{\ \ a}$ is given by

$$F_{(2)}^{\ \ a} = \mu\, u^b\ \nabla_b\, u^c\ \nabla_c\, v^a \tag{9}$$

$$= \mu\, u^b\, v^c\, u^d\, (R_{bcd}^{\ \ \ a} - 2\ \nabla_b\, S_{cd}^{\ \ a}) + 2\,\mu\, u^b\, u^c\, S_{bd}^{\ \ a}\ \nabla_c\, v^d$$

$$= \mu\, u^b\, v^c\, u^d\, (R_{bcd}^{\ \ \ a} - 2\ \nabla_b\, S_{cd}^{\ \ a} - 4\, S_{be}^{\ \ a}\, S_{cd}^{\ \ e}) + 2\,\mu\, u^b\, v^c\, S_{bd}^{\ \ a}\ \nabla_c\, u^d.$$

Finally, calculations of 5-dimensional conservation and geodesic deviation equations appear in Appendix D and formulas for the 5-dimensional force vectors $F_{(1)}^{\ \ a}$ and $F_{(2)}^{\ \ a}$ in Appendix E (see below), the underlying idea being that the extra, "correction" terms appearing therein that are interpretable as thrust determine the structures of electromagnetic or other field configurations (e.g., crossed solenoidal electric and poloidal magnetic) that may—if the configurations are accessible to control—indicate new but practicable means of propulsive force generation.

## Appendix A. Some 5-Dimensional Connection Coefficients and Curvature Tensors

> "There have been some attempts to represent Kaluza's theory formally so as to avoid the introduction of the fifth dimension of the physical continuum. The theory presented here differs from Kaluza's in one essential point; we ascribe physical reality to the fifth dimension whereas in Kaluza's theory this fifth dimension was introduced only in order to obtain new components of the metric tensor representing the electromagnetic field."
>
> — Einstein, A., and P. Bergmann, "On a Generalization of Kaluza's Theory of Electricity," *Annals of Mathematics*, **39** (1938) 683.

> "This theory [i.e., 5-dimensional Kaluza-Klein theory] is surely one of the most remarkable ideas ever advanced for unification of electromagnetism and gravitation."
>
> — Witten, E., "Search for a Realistic Kaluza-Klein Theory," *Nuclear Physics*, **B186** (1981) 412.

Expressions are given below for the Christoffel symbols and connection coefficients of a 5-dimensional differentiable manifold (representing "the physical continuum" comprising space, time, and the 5$^{th}$ dimension) having a non-torsion-free, metrical linear connection accommodating arbitrary functional dependences on the 5$^{th}$ dimension, which expressions are followed by formulas for calculating the corresponding 5-dimensional Riemann-Christoffel, Ricci, and Einstein curvature tensors and Riemann curvature scalar. For alternatives, higher-dimensional differentiable manifolds and associated metric tensors "of the Kaluza-Klein type" incorporating more general[46-56] (e.g., non-Abelian) gauge fields could be used instead.

Greek indices (representing 5-dimensional holonomic coordinates) range from 1 to 5 (1 to 3 corresponding to space, 4 to time, and 5 to the 5$^{th}$ dimension), whereas Latin indices (representing 4-dimensional holonomic coordinates) range from 1 to 4 (1 to 3 corresponding to space and 4 to time).

### 5-Dimensional Co- and Contravariant Metric Tensors

The eight parts of the 5-dimensional co- and contravariant metric tensors ${}^{(5)}g_{\alpha\beta}$ and ${}^{(5)}g^{\alpha\beta}$, respectively, are given by

$${}^{(5)}g_{ab} = g_{ab} + \alpha^2\, A_a\, A_b\, \psi, \tag{10}$$

$${}^{(5)}g_{a5} = \alpha\, A_a\, \psi, \tag{11}$$

$${}^{(5)}g_{5a} = \alpha\, A_a\, \psi, \tag{12}$$

$${}^{(5)}g_{55} = \psi, \tag{13}$$

and

$${}^{(5)}g^{ab} = g^{ab}, \tag{14}$$

$${}^{(5)}g^{a5} = -\,\alpha\, A^a, \tag{15}$$

$${}^{(5)}g^{5a} = -\,\alpha\, A^a, \tag{16}$$

$${}^{(5)}g^{55} = \zeta, \tag{17}$$

respectively, where $g_{ab}$ and $g^{ab}$ are the spacetime co- and contravariant metric tensors, respectively, $\alpha$ is a constant, $A_a$ is the spacetime electromagnetic vector potential, $\psi$ is a scalar, and the scalar $\zeta$ is defined by

$$\zeta = \alpha^2\, \eta + \psi^{-1}, \tag{18}$$

where the scalar $\eta$ is defined by

$$\eta = A_a\, A^a. \tag{19}$$

### 5-Dimensional Christoffel Symbols of the 1$^{st}$ Kind

> "The idea that the electric field quantities are distorted Christoffel symbols has already been haunting me too, often persistently."
>
> — Einstein, A., April 21, 1919, from a letter to T. Kaluza [tr. by author].

The 5-dimensional Christoffel symbols of the 1$^{st}$ kind $[_{\alpha\beta,\gamma}]$ as defined by

$$[_{\alpha\beta,\gamma}] = \frac{1}{2}\,(\partial_\alpha\, {}^{(5)}g_{\beta\gamma} + \partial_\beta\, {}^{(5)}g_{\alpha\gamma} - \partial_\gamma\, {}^{(5)}g_{\alpha\beta}) \tag{20}$$

are given by

$$[_{ab,c}] = -\,\frac{1}{2}\,\alpha^2\,(\partial_c\ \psi)\, A_a\, A_b + \frac{1}{2}\,\alpha^2\,(\partial_b\ \psi)\, A_a\, A_c + \frac{1}{2}\,\alpha^2\,(\partial_a\ \psi)\, A_b\, A_c + \tag{21}$$

$$+ {}^{(4)}[_{ab,c}] + \frac{1}{2}\,\alpha^2\, A_c\, U_{ab}\ \psi + \frac{1}{2}\,\alpha^2\, A_b\, F_{ac}\ \psi + \frac{1}{2}\,\alpha^2\, A_a\, F_{bc}\ \psi,$$

$$[_{ab,5}] = -\,\frac{1}{2}\,\partial_5\, g_{ab} + \frac{1}{2}\,\alpha\,(\partial_b\ \psi)\, A_a + \frac{1}{2}\,\alpha\,(\partial_a\ \psi)\, A_b + \frac{1}{2}\,\alpha\, U_{ab}\ \psi\, - \tag{22}$$

$$-\,\frac{1}{2}\,\alpha^2\,(\partial_5\ \psi)\, A_a\, A_b - \frac{1}{2}\,\alpha^2\,(\partial_5\, A_b)\, A_a\ \psi - \frac{1}{2}\,\alpha^2\,(\partial_5\, A_a)\, A_b\ \psi,$$

$$[_{a5,b}] = \frac{1}{2}\,\partial_5\, g_{ab} - \frac{1}{2}\,\alpha\,(\partial_b\ \psi)\, A_a + \frac{1}{2}\,\alpha\,(\partial_a\ \psi)\, A_b + \frac{1}{2}\,\alpha\, F_{ab}\ \psi\, + \tag{23}$$

$$+\,\frac{1}{2}\,\alpha^2\,(\partial_5\ \psi)\, A_a\, A_b + \frac{1}{2}\,\alpha^2\,(\partial_5\, A_b)\, A_a\ \psi + \frac{1}{2}\,\alpha^2\,(\partial_5\, A_a)\, A_b\ \psi,$$

---

45 Hayden, H. A., "Infinitesimal Deformations of Sub-Spaces in a General Metrical Space," *Proc. London Math. Soc. (2)*, **37** (1934) 416.

46 DeWitt, B. S., "Dynamical Theory of Groups and Fields," in DeWitt, C. M., and B. S. DeWitt (eds.), *Relativity, Groups and Topology, Lectures Delivered at Les Houches During the 1963 Session of the Summer School of Theoretical Physics, University of Grenoble*, Gordon and Breach, Science Publishers, Inc., New York, NY (1964), p. 585, secs. 13 and 16; see also *Dynamical Theory of Groups and Fields*, Gordon and Breach, Science Publishers, Inc., New York, NY (1965), pp. 110 and 138-139.

47 Kerner, R., "Generalizations of the Kaluza-Klein theory for an arbitrary non-abelian [sic] gauge group," *Annales de l'Institut Henri Poincaré, Sect. A (N.S.)*, **9** (1968) 143.

48 Trautmann, A., "Fibre Bundles Associated with Space-Time," *Rep. Math. Phys.*, **1** (1970) 29.

49 Cho, Y. M., and P. G. O. Freund, "Non-Abelian gauge fields as Nambu-Goldstone fields," *Phys. Rev. D*, **12** (1975) 1711.

50 Chang, L. N., K. I. Macrae, and F. Mansouri, "Geometrical approach to local gauge and supergauge invariance: Local gauge theories and supersymmetric strings," *Phys. Rev. D*, **13** (1976) 235.

51 Luciani, J. F., "Space-Time Geometry and Symmetry Breaking," *Nucl. Phys.*, **B135** (1978) 111.

52 Witten, E., "Search for a Realistic Kaluza-Klein Theory," *Nucl. Phys.*, **B186** (1981) 412.

53 Salam, A., and J. Strathdee, "On Kaluza-Klein Theory," *Annals of Physics (N.Y.)*, **141** (1982) 316.

54 Weinberg, S., "Charges from Extra Dimensions," *Phys. Lett.*, **125B** (1983) 265.

55 Overduin, J. M., and P. S. Wesson, "Kaluza-Klein gravity," *Phys. Rep.*, **283** (1997) 303.

56 Vasilić, M., "Higher-dimensional geometric sigma models," *Phys. Rev. D*, **60** (1999) 025003.

$$[_{a5,5}] = \tfrac{1}{2}\,\partial_a\,\psi, \tag{24}$$

$$[_{5a,b}] = \tfrac{1}{2}\,\partial_5\,g_{ab} - \tfrac{1}{2}\,\alpha\,(\partial_b\,\psi)\,A_a + \tfrac{1}{2}\,\alpha\,(\partial_a\,\psi)\,A_b + \tfrac{1}{2}\,\alpha\,F_{ab}\,\psi + \tag{25}$$
$$+\tfrac{1}{2}\,\alpha^2\,(\partial_5\,\psi)\,A_a\,A_b + \tfrac{1}{2}\,\alpha^2\,(\partial_5\,A_b)\,A_a\,\psi + \tfrac{1}{2}\,\alpha^2\,(\partial_5\,A_a)\,A_b\,\psi,$$

$$[_{5a,5}] = \tfrac{1}{2}\,\partial_a\,\psi, \tag{26}$$

$$[_{55,a}] = -\tfrac{1}{2}\,\partial_a\,\psi + \alpha\,(\partial_5\,\psi)\,A_a + \alpha\,(\partial_5\,A_a)\,\psi, \tag{27}$$

$$[_{55,5}] = \tfrac{1}{2}\,\partial_5\,\psi, \tag{28}$$

where the quantities

$$^{(4)}[_{ab,c}] \equiv \tfrac{1}{2}\,(\partial_a\,g_{bc} + \partial_b\,g_{ac} - \partial_c\,g_{ab}) \tag{29}$$

represent the spacetime parts of the 5-dimensional generalization of the usual 4-dimensional spacetime Christoffel symbols of the $2^{nd}$ kind, and the symmetric $2^{nd}$-order spacetime tensor $U_{ab}$ and the anti-symmetric $2^{nd}$-order spacetime tensor $F_{ab}$ are defined by

$$U_{ab} \equiv \partial_a\,A_b + \partial_b\,A_a = 2\,\partial_{(a}\,A_{b)} \tag{30}$$

and

$$F_{ab} \equiv \partial_a\,A_b - \partial_b\,A_a = 2\,\partial_{[a}\,A_{b]}, \tag{31}$$

respectively.

### 5-DIMENSIONAL CHRISTOFFEL SYMBOLS OF THE $2^{nd}$ KIND

The eight parts of the 5-dimensional Christoffel symbols of the $2^{nd}$ kind $\{{}_{\alpha}{}^{\beta}{}_{\gamma}\}$ as defined by

$$\{{}_{\alpha}{}^{\beta}{}_{\gamma}\} = {}^{(5)}g^{\beta\delta}\,[_{\alpha\gamma,\delta}] \tag{32}$$

are given by

$$\{{}_{a}{}^{b}{}_{c}\} = {}^{(5)}g^{bd}\,[_{ac,d}] + {}^{(5)}g^{b5}\,[_{ac,5}] \tag{33}$$
$$= -\,\alpha\,A^b\,[_{ac,5}] + g^{bd}\,[_{ac,d}]$$
$$= \tfrac{1}{2}\,\alpha\,(\partial_5\,g_{ac})\,A^b + \tfrac{1}{2}\,\alpha^3(\partial_5\,\psi)\,A_a\,A_c\,A^b + {}^{(4)}\{{}_{a}{}^{b}{}_{c}\} -$$
$$-\tfrac{1}{2}\,\alpha^2\,(\partial_d\,\psi)\,A_a\,A_c\,g^{bd} + \tfrac{1}{2}\,\alpha^3\,(\partial_5\,A_c)\,A_a\,A^b\,\psi +$$
$$+\tfrac{1}{2}\,\alpha^3\,(\partial_5\,A_a)\,A_c\,A^b\,\psi + \tfrac{1}{2}\,\alpha^2\,A_c\,F_a{}^b\,\psi - \tfrac{1}{2}\,\alpha^2\,A_a\,F^b{}_c\,\psi,$$

$$\{{}_{a}{}^{b}{}_{5}\} = {}^{(5)}g^{bc}\,[_{a5,c}] + {}^{(5)}g^{b5}\,[_{a5,5}] \tag{34}$$
$$= -\,\alpha\,A^b\,[_{a5,5}] + g^{bc}\,[_{a5,c}]$$
$$= \tfrac{1}{2}\,\alpha^2\,(\partial_5\,\psi)\,A_a\,A^b + \tfrac{1}{2}\,(\partial_5\,g_{ac})\,g^{bc} - \tfrac{1}{2}\,\alpha\,(\partial_c\,\psi)\,A_a\,g^{bc} +$$
$$+\tfrac{1}{2}\,\alpha^2\,(\partial_5\,A_a)\,A^b\,\psi + \tfrac{1}{2}\,\alpha\,F_a{}^b\,\psi + \tfrac{1}{2}\,\alpha^2\,(\partial_5\,A_c)\,A_a\,g^{bc}\,\psi,$$

$$\{{}_{a}{}^{5}{}_{b}\} = {}^{(5)}g^{c5}\,[_{ab,c}] + {}^{(5)}g^{55}\,[_{ab,5}] \tag{35}$$
$$= -\,\alpha\,A^c\,[_{ab,c}] + (\alpha^2\,\eta + \psi^{-1})\,[_{ab,5}]$$
$$= -\tfrac{1}{2}\,\alpha^2\,(\partial_5\,A_b)\,A_a - \tfrac{1}{2}\,\alpha^2\,(\partial_5\,A_a)\,A_b + \tfrac{1}{2}\,\alpha^3\,(\partial_c\,\psi)\,A_a\,A_b\,A^c -$$
$$-\,\alpha\,A_c\,{}^{(4)}\{{}_{a}{}^{c}{}_{b}\} + \tfrac{1}{2}\,\alpha\,U_{ab} - \tfrac{1}{2}\,\alpha^2\,(\partial_5\,g_{ab})\,\eta - \tfrac{1}{2}\,\alpha^4\,(\partial_5\,\psi)\,A_a\,A_b\,\eta -$$
$$-\tfrac{1}{2}\,(\partial_5\,g_{ab})\,\psi^{-1} + \tfrac{1}{2}\,\alpha\,(\partial_b\,\psi)\,A_a\,\psi^{-1} + \tfrac{1}{2}\,\alpha\,(\partial_a\,\psi)\,A_b\,\psi^{-1} -$$
$$-\tfrac{1}{2}\,\alpha^2\,(\partial_5\,\psi)\,A_a\,A_b\,\psi^{-1} - \tfrac{1}{2}\,\alpha^4\,(\partial_5\,A_b)\,A_a\,\eta\,\psi -$$
$$-\tfrac{1}{2}\,\alpha^4\,(\partial_5\,A_a)\,A_b\,\eta\,\psi - \tfrac{1}{2}\,\alpha^3\,A_b\,A_c\,F_a{}^c\,\psi - \tfrac{1}{2}\,\alpha^3\,A_a\,A_c\,F_b{}^c\,\psi,$$

$$\{{}_{a}{}^{5}{}_{5}\} = {}^{(5)}g^{b5}\,[_{a5,b}] + {}^{(5)}g^{55}\,[_{a5,5}] \tag{36}$$
$$= -\,\alpha\,A^b\,[_{a5,b}] + (\alpha^2\,\eta + \psi^{-1})\,[_{a5,5}]$$
$$= -\tfrac{1}{2}\,\alpha\,(\partial_5\,g_{ab})\,A^b + \tfrac{1}{2}\,\alpha^2\,(\partial_b\,\psi)\,A_a\,A^b - \tfrac{1}{2}\,\alpha^3\,(\partial_5\,\psi)\,A_a\,\eta +$$
$$+\tfrac{1}{2}\,(\partial_a\,\psi)\,\psi^{-1} - \tfrac{1}{2}\,\alpha^3\,(\partial_5\,A_b)\,A_a\,A^b\,\psi - \tfrac{1}{2}\,\alpha^3\,(\partial_5\,A_a)\,\eta\,\psi -$$
$$-\tfrac{1}{2}\,\alpha^2\,A_b\,F_a{}^b\,\psi,$$

$$\{{}_{5}{}^{a}{}_{b}\} = {}^{(5)}g^{ac}\,[_{b5,c}] + {}^{(5)}g^{a5}\,[_{b5,5}] \tag{37}$$
$$= -\,\alpha\,A^a\,[_{b5,5}] + g^{ac}\,[_{b5,c}]$$
$$= \tfrac{1}{2}\,\alpha^2\,(\partial_5\,\psi)\,A_b\,A^a + \tfrac{1}{2}\,(\partial_5\,g_{bc})\,g^{ac} - \tfrac{1}{2}\,\alpha\,(\partial_c\,\psi)\,A_b\,g^{ac} +$$
$$+\tfrac{1}{2}\,\alpha^2\,(\partial_5\,A_b)\,A^a\,\psi - \tfrac{1}{2}\,\alpha\,F^a{}_b\,\psi + \tfrac{1}{2}\,\alpha^2\,(\partial_5\,A_c)\,A_b\,g^{ac}\,\psi,$$

$$\{{}_{5}{}^{a}{}_{5}\} = {}^{(5)}g^{ab}\,[_{55,b}] + {}^{(5)}g^{a5}\,[_{55,5}] \tag{38}$$
$$= -\,\alpha\,A^a\,[_{55,5}] + g^{ab}\,[_{55,b}]$$
$$= \tfrac{1}{2}\,\alpha\,(\partial_5\,\psi)\,A^a - \tfrac{1}{2}\,(\partial_b\,\psi)\,g^{ab} + \alpha\,(\partial_5\,A_b)\,g^{ab}\,\psi,$$

$$\{{}_{5}{}^{5}{}_{a}\} = {}^{(5)}g^{b5}\,[_{a5,b}] + {}^{(5)}g^{55}\,[_{a5,5}] \tag{39}$$
$$= -\,\alpha\,A^b\,[_{a5,b}] + (\alpha^2\,\eta + \psi^{-1})\,[_{a5,5}]$$
$$= -\tfrac{1}{2}\,\alpha\,(\partial_5\,g_{ab})\,A^b + \tfrac{1}{2}\,\alpha^2\,(\partial_b\,\psi)\,A_a\,A^b - \tfrac{1}{2}\,\alpha^3\,(\partial_5\,\psi)\,A_a\,\eta +$$
$$+\tfrac{1}{2}\,(\partial_a\,\psi)\,\psi^{-1} - \tfrac{1}{2}\,\alpha^3\,(\partial_5\,A_b)\,A_a\,A^b\,\psi - \tfrac{1}{2}\,\alpha^3\,(\partial_5\,A_a)\,\eta\,\psi -$$
$$-\tfrac{1}{2}\,\alpha^2\,A_b\,F_a{}^b\,\psi,$$

$$\{{}_{5}{}^{5}{}_{5}\} = {}^{(5)}g^{a5}\,[_{55,a}] + {}^{(5)}g^{55}\,[_{55,5}] \tag{40}$$
$$= -\,\alpha\,A^a\,[_{55,a}] + (\alpha^2\,\eta + \psi^{-1})\,[_{55,5}]$$
$$= \tfrac{1}{2}\,\alpha\,(\partial_a\,\psi)\,A^a - \tfrac{1}{2}\,\alpha^2\,(\partial_5\,\psi)\,\eta + \tfrac{1}{2}\,(\partial_5\,\psi)\,\psi^{-1} - \alpha^2\,(\partial_5\,A_a)\,A^a\,\psi,$$

where

$$^{(4)}\{{}_{a}{}^{b}{}_{c}\} \equiv g^{bd}\,{}^{(4)}[_{ac,d}] = \tfrac{1}{2}\,g^{bd}\,(\partial_a\,g_{cd} + \partial_c\,g_{ad} - \partial_d\,g_{ac}). \tag{41}$$

### 5-DIMENSIONAL TORSION TENSOR

The eight parts of the 5-dimensional torsion tensor $S_{\alpha\beta}{}^{\gamma}$ are given by

$$S_{ab}{}^{c} = {}^{(4)}S_{ab}{}^{c}, \tag{42}$$

$$S_{ab}{}^{5} = \beta_{(1)}\,S_{(1)ab}, \tag{43}$$

$$S_{a5}{}^{b} = \beta_{(2)}\,S_{(2)a}{}^{b}, \tag{44}$$

$$S_{a5}{}^{5} = \beta_{(3)}\,S_{(3)a}, \tag{45}$$

$$S_{5a}{}^{b} = -\,\beta_{(2)}\,S_{(2)a}{}^{b}, \tag{46}$$

$$S_{5a}{}^{5} = -\,\beta_{(3)}\,S_{(3)a}, \tag{47}$$

$$S_{55}{}^{a} = 0, \tag{48}$$

$$S_{55}{}^{5} = 0, \tag{49}$$

where $\beta_{(1)}$, $\beta_{(2)}$, and $\beta_{(3)}$ are constants, ${}^{(4)}S_{ab}{}^{c}$ is the spacetime part of the torsion tensor, $S_{(1)ab} = -\,S_{(1)ba}$ and $S_{(2)a}{}^{b} = g^{bc}\,S_{(2)ac}$ are $2^{nd}$-order spacetime tensors, and $S_{(3)a}$ is a spacetime vector.

### 5-DIMENSIONAL CONNECTION COEFFICIENTS

The eight parts of the 5-dimensional connection coefficients $\Gamma_{\alpha}{}^{\beta}{}_{\gamma}$ as defined (using holonomic coordinates for a non-torsion-free, metrical linear connection) by[57]

$$\Gamma_{\alpha}{}^{\beta}{}_{\gamma} = \{{}_{\alpha}{}^{\beta}{}_{\gamma}\} + S_{\alpha\gamma}{}^{\beta} - S_{\alpha}{}^{\beta}{}_{\gamma} + S^{\beta}{}_{\gamma\alpha} = \{{}_{\alpha}{}^{\beta}{}_{\gamma}\} + S_{\alpha\gamma}{}^{\beta} - S_{\gamma}{}^{\beta}{}_{\alpha} + S^{\beta}{}_{\alpha\gamma} \tag{50}$$

are given by

$$\Gamma_{a}{}^{b}{}_{c} = \tfrac{1}{2}\,\alpha\,(\partial_5\,g_{ac})\,A^b + \tfrac{1}{2}\,\alpha^3(\partial_5\,\psi)\,A_a\,A_c\,A^b - \tfrac{1}{2}\,\alpha^2\,(\partial_d\,\psi)\,A_a\,A_c\,g^{bd} + \tag{51}$$
$$+\,{}^{(4)}\{{}_{a}{}^{b}{}_{c}\} + \tfrac{1}{2}\,\alpha^3\,(\partial_5\,A_c)\,A_a\,A^b\,\psi + \tfrac{1}{2}\,\alpha^3\,(\partial_5\,A_a)\,A_c\,A^b\,\psi +$$
$$+\tfrac{1}{2}\,\alpha^2\,A_c\,F_a{}^b\,\psi - \tfrac{1}{2}\,\alpha^2\,A_a\,F^b{}_c\,\psi + {}^{(4)}S_{ac}{}^{b} - {}^{(4)}S_{a}{}^{b}{}_{c} + {}^{(4)}S^{b}{}_{ca} -$$
$$-\,\alpha^2\,A_c\,A_d\,{}^{(4)}S_{a}{}^{bd}\,\psi + \alpha^2\,A_a\,A_d\,{}^{(4)}S^{b}{}_{c}{}^{d}\,\psi - \alpha\,\beta_{(1)}\,A_c\,\psi\,S_{(1)a}{}^{b} +$$
$$+\,\alpha\,\beta_{(1)}\,A_a\,\psi\,S_{(1)}{}^{b}{}_{c} + \alpha^3\,\beta_{(2)}\,A_c\,A_d\,A^b\,\psi\,S_{(2)a}{}^{d} +$$

[57] Schouten, J. A., *Ricci-Calculus*, Springer-Verlag, Berlin, Germany (1954), p. 132; cf. p. 170.

$$\begin{aligned}
&+ \alpha^3\, \beta_{(2)}\, A_a\, A_d\, A^b\, \psi\, S_{(2)c}{}^{d} + \alpha^2\, \beta_{(3)}\, A_c\, A^b\, \psi\, S_{(3)a} + \\
&+ \alpha^2\, \beta_{(3)}\, A_a\, A^b\, \psi\, S_{(3)c},
\end{aligned}$$

$$\begin{aligned}
\Gamma_{a}{}^{b}{}_{5} = {} & \tfrac{1}{2}\, \alpha^2\, (\partial_5\, \psi)\, A_a\, A^b + \tfrac{1}{2}\, (\partial_5\, g_{ac})\, g^{bc} - \tfrac{1}{2}\, \alpha\, (\partial_c\, \psi)\, A_a\, g^{bc} + \\
&+ \tfrac{1}{2}\, \alpha^2\, (\partial_5\, A_a)\, A^b\, \psi + \tfrac{1}{2}\, \alpha\, F_a{}^{b}\, \psi + \tfrac{1}{2}\, \alpha^2\, (\partial_5\, A_c)\, A_a\, g^{bc}\, \psi - \\
&- \alpha\, A_c\, {}^{(4)}S_a{}^{bc}\, \psi - \beta_{(1)}\, \psi\, S_{(1)a}{}^{b} + \alpha^2\, \beta_{(2)}\, A_c\, A^b\, \psi\, S_{(2)a}{}^{c} + \\
&+ \alpha^2\, \beta_{(2)}\, A_a\, A_c\, \psi\, S_{(2)}{}^{bc} + \alpha\, \beta_{(3)}\, A^b\, \psi\, S_{(3)a} + \alpha\, \beta_{(3)}\, A_a\, \psi\, S_{(3)}{}^{b},
\end{aligned} \tag{52}$$

$$\begin{aligned}
\Gamma_{a}{}^{5}{}_{b} = {} & -\tfrac{1}{2}\, \alpha^2\, (\partial_5\, A_b)\, A_a - \tfrac{1}{2}\, \alpha^2\, (\partial_5\, A_a)\, A_b + \tfrac{1}{2}\, \alpha^3\, (\partial_c\, \psi)\, A_a\, A_b\, A^c - \\
&- \alpha\, A_c\, {}^{(4)}\{{}^{c}_{ab}\} + \tfrac{1}{2}\, \alpha\, U_{ab} - \tfrac{1}{2}\, \alpha^2\, (\partial_5\, g_{ab})\, \eta - \tfrac{1}{2}\, \alpha^4\, (\partial_5\, \psi)\, A_a\, A_b\, \eta - \\
&- \tfrac{1}{2}\, (\partial_5\, g_{ab})\, \psi^{-1} + (\partial_b\, \psi)\, A_a\, \psi^{-1} + \tfrac{1}{2}\, \alpha\, (\partial_a\, \psi)\, A_b\, \psi^{-1} - \\
&- \tfrac{1}{2}\, \alpha^2\, (\partial_5\, \psi)\, A_a\, A_b\, \psi^{-1} - \tfrac{1}{2}\, \alpha^4\, (\partial_5\, A_b)\, A_a\, \eta\, \psi - \\
&- \tfrac{1}{2}\, \alpha^4\, (\partial_5\, A_a)\, A_b\, \eta\, \psi - \tfrac{1}{2}\, \alpha^3\, A_b\, A_c\, F_a{}^{c}\, \psi - \tfrac{1}{2}\, \alpha^3\, A_a\, A_c\, F_b{}^{c}\, \psi + \\
&+ \alpha\, A_c\, {}^{(4)}S_a{}^{c}{}_{b} + \alpha\, A_c\, {}^{(4)}S_b{}^{c}{}_{a} + \alpha^3\, A_b\, A_c\, A_d\, {}^{(4)}S_a{}^{cd}\, \psi + \\
&+ \alpha^3\, A_a\, A_c\, A_d\, {}^{(4)}S_b{}^{cd}\, \psi + \beta_{(1)}\, \psi\, S_{(1)ab} + \alpha^2\, \beta_{(1)}\, A_b\, A_c\, \psi\, S_{(1)a}{}^{c} + \\
&+ \alpha^2\, \beta_{(1)}\, A_a\, A_c\, \psi\, S_{(1)b}{}^{c} - \alpha^2\, \beta_{(2)}\, A_b\, A_c\, S_{(2)a}{}^{c} - \\
&- \alpha^4\, \beta_{(2)}\, A_b\, A_c\, \eta\, \psi\, S_{(2)a}{}^{c} - \alpha^2\, \beta_{(2)}\, A_a\, A_c\, S_{(2)b}{}^{c} - \\
&- \alpha^4\, \beta_{(2)}\, A_a\, A_c\, \eta\, \psi\, S_{(2)b}{}^{c} - \alpha\, \beta_{(3)}\, A_b\, S_{(3)a} - \\
&- \alpha^3\, \beta_{(3)}\, A_b\, \eta\, \psi\, S_{(3)a} - \alpha\, \beta_{(3)}\, A_a\, S_{(3)b} - \alpha^3\, \beta_{(3)}\, A_a\, \eta\, \psi\, S_{(3)b},
\end{aligned} \tag{53}$$

$$\begin{aligned}
\Gamma_{a}{}^{5}{}_{5} = {} & -\tfrac{1}{2}\, \alpha\, (\partial_5\, g_{ab})\, A^b + \tfrac{1}{2}\, \alpha^2\, (\partial_b\, \psi)\, A_a\, A^b - \tfrac{1}{2}\, \alpha^3\, (\partial_5\, \psi)\, A_a\, \eta + \\
&+ \tfrac{1}{2}\, (\partial_a\, \psi)\, \psi^{-1} - \tfrac{1}{2}\, \alpha^3\, (\partial_5\, A_b)\, A_a\, A^b\, \psi + \alpha^2\, A_b\, A_c\, {}^{(4)}S_a{}^{bc}\, \psi - \\
&- \tfrac{1}{2}\, \alpha^3\, (\partial_5\, A_a)\, \eta\, \psi - \tfrac{1}{2}\, \alpha^2\, A_b\, F_a{}^{b}\, \psi + \alpha\, \beta_{(1)}\, A_b\, \psi\, S_{(1)a}{}^{b} - \\
&- \alpha^3\, \beta_{(2)}\, A_b\, \eta\, \psi\, S_{(2)a}{}^{b} - \alpha^2\, \beta_{(3)}\, \eta\, \psi\, S_{(3)a} - \alpha^2\, \beta_{(3)}\, A_a\, A_b\, \psi\, S_{(3)}{}^{b},
\end{aligned} \tag{54}$$

$$\begin{aligned}
\Gamma_{5}{}^{a}{}_{b} = {} & \tfrac{1}{2}\, \alpha^2\, (\partial_5\, \psi)\, A_b\, A^a + \tfrac{1}{2}\, (\partial_5\, g_{bc})\, g^{ac} - \tfrac{1}{2}\, \alpha\, (\partial_c\, \psi)\, A_b\, g^{ac} + \\
&+ \tfrac{1}{2}\, \alpha^2\, (\partial_5\, A_b)\, A^a\, \psi - \tfrac{1}{2}\, \alpha\, F^a{}_{b}\, \psi + \tfrac{1}{2}\, \alpha^2\, (\partial_5\, A_c)\, A_b\, g^{ac}\, \psi + \\
&+ \alpha\, A_c\, {}^{(4)}S^a{}_{b}{}^{c}\, \psi + \beta_{(1)}\, \psi\, S_{(1)}{}^{a}{}_{b} + \alpha^2\, \beta_{(2)}\, A_c\, A^a\, \psi\, S_{(2)b}{}^{c} + \\
&+ 2\, \beta_{(2)}\, S_{(2)}{}^{a}{}_{b} + \alpha^2\, \beta_{(2)}\, A_b\, A_c\, \psi\, S_{(2)}{}^{ac} + \alpha\, \beta_{(3)}\, A^a\, \psi\, S_{(3)b} + \\
&+ \alpha\, \beta_{(3)}\, A_b\, \psi\, S_{(3)}{}^{a},
\end{aligned} \tag{55}$$

$$\begin{aligned}
\Gamma_{5}{}^{a}{}_{5} = {} & \tfrac{1}{2}\, \alpha\, (\partial_5\, \psi)\, A^a - \tfrac{1}{2}\, (\partial_b\, \psi)\, g^{ab} + \alpha\, (\partial_5\, A_b)\, g^{ab}\, \psi + \\
&+ 2\, \alpha\, \beta_{(2)}\, A_b\, \psi\, S_{(2)}{}^{ab} + 2\, \beta_{(3)}\, \psi\, S_{(3)}{}^{a},
\end{aligned} \tag{56}$$

$$\begin{aligned}
\Gamma_{5}{}^{5}{}_{a} = {} & -\tfrac{1}{2}\, \alpha\, (\partial_5\, g_{ab})\, A^b + \tfrac{1}{2}\, \alpha^2\, (\partial_b\, \psi)\, A_a\, A^b - \tfrac{1}{2}\, \alpha^3\, (\partial_5\, \psi)\, A_a\, \eta + \\
&+ \tfrac{1}{2}\, (\partial_a\, \psi)\, \psi^{-1} - \tfrac{1}{2}\, \alpha^3\, (\partial_5\, A_b)\, A_a\, A^b\, \psi - \tfrac{1}{2}\, \alpha^3\, (\partial_5\, A_a)\, \eta\, \psi - \\
&- \tfrac{1}{2}\, \alpha^2\, A_b\, F_a{}^{b}\, \psi + \alpha^2\, A_b\, A_c\, {}^{(4)}S_a{}^{bc}\, \psi + \alpha\, \beta_{(1)}\, A_b\, \psi\, S_{(1)a}{}^{b} - \\
&- \alpha^3\, \beta_{(2)}\, A_b\, \eta\, \psi\, S_{(2)a}{}^{b} - 2\, \beta_{(3)}\, S_{(3)a} - \alpha^2\, \beta_{(3)}\, \eta\, \psi\, S_{(3)a} - \\
&- \alpha^2\, \beta_{(3)}\, A_a\, A_b\, \psi\, S_{(3)}{}^{b},
\end{aligned} \tag{57}$$

$$\begin{aligned}
\Gamma_{5}{}^{5}{}_{5} = {} & \tfrac{1}{2}\, \alpha\, (\partial_a\, \psi)\, A^a - \tfrac{1}{2}\, \alpha^2\, (\partial_5\, \psi)\, \eta + \tfrac{1}{2}\, (\partial_5\, \psi)\, \psi^{-1} - \\
&- \alpha^2\, (\partial_5\, A_a)\, A^a\, \psi - 2\, \alpha\, \beta_{(3)}\, A_a\, \psi\, S_{(3)}{}^{a}.
\end{aligned} \tag{58}$$

### 5-DIMENSIONAL RIEMANN-CHRISTOFFEL CURVATURE TENSOR

The 16 parts of the 5-dimensional Riemann-Christoffel curvature tensor $R_{\alpha\beta\gamma}{}^{\delta}$ as defined (using holonomic coordinates for a non-torsion-free, metrical linear connection) by[58]

$$R_{\alpha\beta\gamma}{}^{\delta} = 2\, (\partial_{[\alpha}\, \Gamma_{\beta]}{}^{\delta}{}_{\gamma} + \Gamma_{[\alpha}{}^{\delta}{}_{|\varepsilon|}\, \Gamma_{\beta]}{}^{\varepsilon}{}_{\gamma}) \tag{59}$$

are given by

$$R_{abc}{}^{d} = 2\, (\partial_{[a}\, \Gamma_{b]}{}^{d}{}_{c} + \Gamma_{[a}{}^{d}{}_{|e|}\, \Gamma_{b]}{}^{e}{}_{c} + \Gamma_{[a}{}^{d}{}_{|5|}\, \Gamma_{b]}{}^{5}{}_{c}), \tag{60}$$

$$R_{abc}{}^{5} = 2\, (\partial_{[a}\, \Gamma_{b]}{}^{5}{}_{c} + \Gamma_{[a}{}^{5}{}_{|d|}\, \Gamma_{b]}{}^{d}{}_{c} + \Gamma_{[a}{}^{5}{}_{|5|}\, \Gamma_{b]}{}^{5}{}_{c}), \tag{61}$$

$$R_{ab5}{}^{c} = 2\, (\partial_{[a}\, \Gamma_{b]}{}^{c}{}_{5} + \Gamma_{[a}{}^{c}{}_{|d|}\, \Gamma_{b]}{}^{d}{}_{5} + \Gamma_{[a}{}^{c}{}_{|5|}\, \Gamma_{b]}{}^{5}{}_{5}), \tag{62}$$

$$\begin{aligned}
R_{ab5}{}^{5} &= 2\, (\partial_{[a}\, \Gamma_{b]}{}^{5}{}_{5} + \Gamma_{[a}{}^{5}{}_{|c|}\, \Gamma_{b]}{}^{c}{}_{5} + \Gamma_{[a}{}^{5}{}_{|5|}\, \Gamma_{b]}{}^{5}{}_{5}) \\
&= 2\, (\partial_{[a}\, \Gamma_{b]}{}^{5}{}_{5} + \Gamma_{[a}{}^{5}{}_{|c|}\, \Gamma_{b]}{}^{c}{}_{5}),
\end{aligned} \tag{63}$$

$$R_{a5b}{}^{c} = 2\, (\partial_{[a}\, \Gamma_{5]}{}^{c}{}_{b} + \Gamma_{[a}{}^{c}{}_{|d|}\, \Gamma_{5]}{}^{d}{}_{b} + \Gamma_{[a}{}^{c}{}_{|5|}\, \Gamma_{5]}{}^{5}{}_{b}), \tag{64}$$

$$R_{a5b}{}^{5} = 2\, (\partial_{[a}\, \Gamma_{5]}{}^{5}{}_{b} + \Gamma_{[a}{}^{5}{}_{|c|}\, \Gamma_{5]}{}^{c}{}_{b} + \Gamma_{[a}{}^{5}{}_{|5|}\, \Gamma_{5]}{}^{5}{}_{b}), \tag{65}$$

$$R_{a55}{}^{b} = 2\, (\partial_{[a}\, \Gamma_{5]}{}^{b}{}_{5} + \Gamma_{[a}{}^{b}{}_{|c|}\, \Gamma_{5]}{}^{c}{}_{5} + \Gamma_{[a}{}^{b}{}_{|5|}\, \Gamma_{5]}{}^{5}{}_{5}), \tag{66}$$

$$\begin{aligned}
R_{a55}{}^{5} &= 2\, (\partial_{[a}\, \Gamma_{5]}{}^{5}{}_{5} + \Gamma_{[a}{}^{5}{}_{|b|}\, \Gamma_{5]}{}^{b}{}_{5} + \Gamma_{[a}{}^{5}{}_{|5|}\, \Gamma_{5]}{}^{5}{}_{5}) \\
&= 2\, (\partial_{[a}\, \Gamma_{5]}{}^{5}{}_{5} + \Gamma_{[a}{}^{5}{}_{|b|}\, \Gamma_{5]}{}^{b}{}_{5}),
\end{aligned} \tag{67}$$

$$R_{5ab}{}^{c} = 2\, (\partial_{[5}\, \Gamma_{a]}{}^{c}{}_{b} + \Gamma_{[5}{}^{c}{}_{|d|}\, \Gamma_{a]}{}^{d}{}_{b} + \Gamma_{[5}{}^{c}{}_{|5|}\, \Gamma_{a]}{}^{5}{}_{b}), \tag{68}$$

$$R_{5ab}{}^{5} = 2\, (\partial_{[5}\, \Gamma_{a]}{}^{5}{}_{b} + \Gamma_{[5}{}^{5}{}_{|c|}\, \Gamma_{a]}{}^{c}{}_{b} + \Gamma_{[5}{}^{5}{}_{|5|}\, \Gamma_{a]}{}^{5}{}_{b}), \tag{69}$$

$$R_{5a5}{}^{b} = 2\, (\partial_{[5}\, \Gamma_{a]}{}^{b}{}_{5} + \Gamma_{[5}{}^{b}{}_{|c|}\, \Gamma_{a]}{}^{c}{}_{5} + \Gamma_{[5}{}^{b}{}_{|5|}\, \Gamma_{a]}{}^{5}{}_{5}), \tag{70}$$

$$\begin{aligned}
R_{5a5}{}^{5} &= 2\, (\partial_{[5}\, \Gamma_{a]}{}^{5}{}_{5} + \Gamma_{[5}{}^{5}{}_{|b|}\, \Gamma_{a]}{}^{b}{}_{5} + \Gamma_{[5}{}^{5}{}_{|5|}\, \Gamma_{a]}{}^{5}{}_{5}) \\
&= 2\, (\partial_{[5}\, \Gamma_{a]}{}^{5}{}_{5} + \Gamma_{[5}{}^{5}{}_{|b|}\, \Gamma_{a]}{}^{b}{}_{5}),
\end{aligned} \tag{71}$$

$$\begin{aligned}
R_{55a}{}^{b} &= 2\, (\partial_{[5}\, \Gamma_{5]}{}^{b}{}_{a} + \Gamma_{[5}{}^{b}{}_{|c|}\, \Gamma_{5]}{}^{c}{}_{a} + \Gamma_{[5}{}^{b}{}_{|5|}\, \Gamma_{5]}{}^{5}{}_{a}) \\
&= 0,
\end{aligned} \tag{72}$$

$$\begin{aligned}
R_{55a}{}^{5} &= 2\, (\partial_{[5}\, \Gamma_{5]}{}^{5}{}_{a} + \Gamma_{[5}{}^{5}{}_{|b|}\, \Gamma_{5]}{}^{b}{}_{a} + \Gamma_{[5}{}^{5}{}_{|5|}\, \Gamma_{5]}{}^{5}{}_{a}) \\
&= 0,
\end{aligned} \tag{73}$$

$$\begin{aligned}
R_{555}{}^{a} &= 2\, (\partial_{[5}\, \Gamma_{5]}{}^{a}{}_{5} + \Gamma_{[5}{}^{a}{}_{|b|}\, \Gamma_{5]}{}^{b}{}_{5} + \Gamma_{[5}{}^{a}{}_{|5|}\, \Gamma_{5]}{}^{5}{}_{5}) \\
&= 0,
\end{aligned} \tag{74}$$

$$\begin{aligned}
R_{555}{}^{5} &= 2\, (\partial_{[5}\, \Gamma_{5]}{}^{5}{}_{5} + \Gamma_{[5}{}^{5}{}_{|b|}\, \Gamma_{5]}{}^{b}{}_{5} + \Gamma_{[5}{}^{5}{}_{|5|}\, \Gamma_{5]}{}^{5}{}_{5}) \\
&= 0.
\end{aligned} \tag{75}$$

### 5-DIMENSIONAL RICCI CURVATURE TENSOR

The four parts of the 5-dimensional Ricci curvature tensor $R_{\alpha\beta}$ as defined (using holonomic coordinates for a general linear connection) by

$$R_{\alpha\beta} = R_{\gamma\alpha\beta}{}^{\gamma} \tag{76}$$

are given by

$$\begin{aligned}
R_{ab} &= R_{cab}{}^{c} + R_{5ab}{}^{5} \\
&= 2\, (\partial_{[c}\, \Gamma_{a]}{}^{c}{}_{b} + \Gamma_{[c}{}^{c}{}_{|d|}\, \Gamma_{a]}{}^{d}{}_{b} + \Gamma_{[c}{}^{c}{}_{|5|}\, \Gamma_{a]}{}^{5}{}_{b}) + \\
&\quad + 2\, (\partial_{[5}\, \Gamma_{a]}{}^{5}{}_{b} + \Gamma_{[5}{}^{5}{}_{|c|}\, \Gamma_{a]}{}^{c}{}_{b} + \Gamma_{[5}{}^{5}{}_{|5|}\, \Gamma_{a]}{}^{5}{}_{b}),
\end{aligned} \tag{77}$$

$$\begin{aligned}
R_{a5} &= R_{ba5}{}^{b} + R_{5a5}{}^{5} \\
&= 2\, (\partial_{[b}\, \Gamma_{a]}{}^{b}{}_{5} + \Gamma_{[b}{}^{b}{}_{|c|}\, \Gamma_{a]}{}^{c}{}_{5} + \Gamma_{[b}{}^{b}{}_{|5|}\, \Gamma_{a]}{}^{5}{}_{5}) + \\
&\quad + 2\, (\partial_{[5}\, \Gamma_{a]}{}^{5}{}_{5} + \Gamma_{[5}{}^{5}{}_{|b|}\, \Gamma_{a]}{}^{b}{}_{5}),
\end{aligned} \tag{78}$$

$$\begin{aligned}
R_{5a} &= R_{b5a}{}^{b} + R_{55a}{}^{5} \\
&= 2\, (\partial_{[b}\, \Gamma_{5]}{}^{b}{}_{a} + \Gamma_{[b}{}^{b}{}_{|c|}\, \Gamma_{5]}{}^{c}{}_{a} + \Gamma_{[b}{}^{b}{}_{|5|}\, \Gamma_{5]}{}^{5}{}_{a}),
\end{aligned} \tag{79}$$

$$\begin{aligned}
R_{55} &= R_{a55}{}^{a} + R_{555}{}^{5} \\
&= 2\, (\partial_{[a}\, \Gamma_{5]}{}^{a}{}_{5} + \Gamma_{[a}{}^{a}{}_{|b|}\, \Gamma_{5]}{}^{b}{}_{5} + \Gamma_{[a}{}^{a}{}_{|5|}\, \Gamma_{5]}{}^{5}{}_{5}).
\end{aligned} \tag{80}$$

### 5-DIMENSIONAL RIEMANN CURVATURE SCALAR

The 5-dimensional Riemann curvature scalar $R$ as defined (using anholonomic coordinates for a general linear connection) by

[58] Schouten, J. A., *ibid.*, p. 138; cf. p. 172.

$$R = R_{\alpha}{}^{\alpha} \tag{81}$$
$$= R_{\beta\alpha}{}^{\alpha\beta}$$

is given by

$$R = R_{\alpha}{}^{\alpha} \tag{82}$$
$$= R_{\beta\alpha}{}^{\alpha\beta}$$
$$= {}^{(5)}g^{\alpha\beta}\, R_{\alpha\beta}$$
$$= {}^{(5)}g^{ab}\, R_{ab} + {}^{(5)}g^{a5}\, R_{a5} + {}^{(5)}g^{5a}\, R_{5a} + {}^{(5)}g^{55}\, R_{55}$$
$$= {}^{(5)}g^{ab}\, R_{ab} + {}^{(5)}g^{a5}\, (R_{a5} + R_{5a}) + {}^{(5)}g^{55}\, R_{55}.$$

5-DIMENSIONAL EINSTEIN CURVATURE TENSOR

The four parts of the 5-dimensional Einstein curvature tensor $G_{\alpha\beta}$ as defined (using anholonomic coordinates for a general linear connection) by

$$G_{\alpha\beta} = R_{\alpha\beta} - \tfrac{1}{2}\, {}^{(5)}g_{\alpha\beta},\, R \tag{83}$$

are given by

$$G_{ab} = R_{ab} - \tfrac{1}{2}\, {}^{(5)}g_{ab}\, R, \tag{84}$$
$$G_{a5} = R_{a5} - \tfrac{1}{2}\, {}^{(5)}g_{a5}\, R, \tag{85}$$
$$G_{5a} = R_{5a} - \tfrac{1}{2}\, {}^{(5)}g_{5a}\, R, \tag{86}$$
$$G_{55} = R_{55} - \tfrac{1}{2}\, {}^{(5)}g_{55}\, R. \tag{87}$$

## APPENDIX B. CALCULATION OF *n*-DIMENSIONAL IDENTITIES FOR THE EINSTEIN CURVATURE TENSOR

Bianchi's $5^{th}$-order identities for the Riemann-Christoffel curvature tensor $R_{abc}{}^{d}$ as defined (using anholonomic coordinates for a general linear connection) by[59]

$$R_{abc}{}^{d} = 2\, (\partial_{[a}\, \Gamma_{b]}{}^{d}{}_{c} + \Gamma_{[a}{}^{d}{}_{|e|}\, \Gamma_{b]}{}^{e}{}_{c} + \Omega_{a}{}^{e}{}_{b}\, \Gamma_{e}{}^{d}{}_{c}), \tag{88}$$

where $\Omega_{a}{}^{b}{}_{c}$ is the object of anholonomity (which vanishes for holonomic coordinates), are given by[60]

$$\nabla_{[a}\, R_{bc]d}{}^{e} = 2\, S_{[ab}{}^{f}\, R_{c]fd}{}^{e}. \tag{89}$$

One $3^{rd}$-order $1^{st}$ contraction of Eq. (89) is given by

$$g^{ad}\, \nabla_{[a}\, R_{bc]d}{}^{e} = \nabla_{[a}\, g^{ad}\, R_{bc]d}{}^{e} - Q_{[a}{}^{ad}\, R_{bc]d}{}^{e} \tag{90}$$
$$= \nabla_{[a}\, R_{bc]}{}^{ae} - Q_{[a}{}^{ad}\, R_{bc]d}{}^{e}$$
$$= \nabla_{[a}\, [- R_{bc]}{}^{ea} + 2\, (\nabla_{b}\, Q_{c]}{}^{ae} + S_{bc]}{}^{d}\, Q_{d}{}^{ae})] - Q_{[a}{}^{ad}\, R_{bc]d}{}^{e}$$
$$= - \nabla_{[a}\, R_{bc]}{}^{ea} + 2\, \nabla_{[a}\, (\nabla_{b}\, Q_{c]}{}^{ae} + S_{bc]}{}^{d}\, Q_{d}{}^{ae}) - Q_{[a}{}^{ad}\, R_{bc]d}{}^{e}$$
$$= - \tfrac{1}{3}\, (\nabla_{a}\, R_{bc}{}^{ea} + \nabla_{b}\, R_{ca}{}^{ea} + \nabla_{c}\, R_{ab}{}^{ea}) +$$
$$+ 2\, \nabla_{[a}\, (\nabla_{b}\, Q_{c]}{}^{ae} + S_{bc]}{}^{d}\, Q_{d}{}^{ae}) - Q_{[a}{}^{ad}\, R_{bc]d}{}^{e}$$
$$= - \tfrac{1}{3}\, (\nabla_{a}\, R_{bc}{}^{ea} - 2\, \nabla_{[b}\, R_{c]}{}^{e}) +$$
$$+ 2\, \nabla_{[a}\, (\nabla_{b}\, Q_{c]}{}^{ae} + S_{bc]}{}^{d}\, Q_{d}{}^{ae}) - Q_{[a}{}^{ad}\, R_{bc]d}{}^{e}$$
$$= 2\, S_{[ab}{}^{f}\, R_{c]f}{}^{ae}$$
$$= \tfrac{2}{3}\, (S_{ab}{}^{f}\, R_{cf}{}^{ae} + S_{bc}{}^{f}\, R_{af}{}^{ae} + S_{ca}{}^{f}\, R_{bf}{}^{ae})$$
$$= \tfrac{2}{3}\, (- S_{ab}{}^{f}\, R_{cf}{}^{ea} - S_{bc}{}^{f}\, R_{af}{}^{ea} - S_{ca}{}^{f}\, R_{bf}{}^{ea}) +$$
$$+ \tfrac{4}{3}\, [S_{ab}{}^{f}\, (\nabla_{[c}\, Q_{f]}{}^{ae} + S_{cf}{}^{d}\, Q_{d}{}^{ae}) +$$
$$+ S_{bc}{}^{f}\, (\nabla_{[a}\, Q_{f]}{}^{ae} + S_{af}{}^{d}\, Q_{d}{}^{ae}) +$$
$$+ S_{ca}{}^{f}\, (\nabla_{[b}\, Q_{f]}{}^{ae} + S_{bf}{}^{d}\, Q_{d}{}^{ae})]$$
$$= - \tfrac{2}{3}\, (2\, S_{a[b}{}^{f}\, R_{c]f}{}^{ea} + S_{bc}{}^{f}\, R_{f}{}^{e}) + 4\, S_{[ab}{}^{f}\, S_{c]f}{}^{d}\, Q_{d}{}^{ae} +$$
$$+ \tfrac{4}{3}\, (S_{ab}{}^{f}\, \nabla_{[c}\, Q_{f]}{}^{ae} + S_{bc}{}^{f}\, \nabla_{[a}\, Q_{f]}{}^{ae} + S_{ca}{}^{f}\, \nabla_{[b}\, Q_{f]}{}^{ae})$$
$$= - \tfrac{2}{3}\, (2\, S_{a[b}{}^{f}\, R_{c]f}{}^{ea} + S_{bc}{}^{f}\, R_{f}{}^{e}) + 4\, S_{[ab}{}^{f}\, S_{c]f}{}^{d}\, Q_{d}{}^{ae} +$$
$$+ \tfrac{2}{3}\, [S_{ab}{}^{f}\, (\nabla_{c}\, Q_{f}{}^{ae} - \nabla_{f}\, Q_{c}{}^{ae}) + S_{bc}{}^{f}\, (\nabla_{a}\, Q_{f}{}^{ae} - \nabla_{f}\, Q_{a}{}^{ae}) +$$
$$+ S_{ca}{}^{f}\, (\nabla_{b}\, Q_{f}{}^{ae} - \nabla_{f}\, Q_{b}{}^{ae})]$$
$$= - \tfrac{2}{3}\, (2\, S_{a[b}{}^{f}\, R_{c]f}{}^{ea} + S_{bc}{}^{f}\, R_{f}{}^{e}) + 4\, S_{[ab}{}^{f}\, S_{c]f}{}^{d}\, Q_{d}{}^{ae} +$$
$$+ 2\, (S_{[ab}{}^{f}\, \nabla_{c]}\, Q_{f}{}^{ae} - S_{[ab}{}^{f}\, \nabla_{|f|}\, Q_{c]}{}^{ae})$$
$$= - \tfrac{2}{3}\, (2\, S_{a[b}{}^{f}\, R_{c]f}{}^{ea} + S_{bc}{}^{f}\, R_{f}{}^{e}) + 4\, S_{[ab}{}^{f}\, S_{c]f}{}^{d}\, Q_{d}{}^{ae} +$$
$$+ 2\, (\nabla_{[a}\, S_{bc]}{}^{f}\, Q_{f}{}^{ae} - Q_{f}{}^{ae}\, \nabla_{[a}\, S_{bc]}{}^{f} - S_{[ab}{}^{f}\, \nabla_{|f|}\, Q_{c]}{}^{ae}),$$

one $1^{st}$-order $2^{nd}$ contraction of which is given by

$$\delta^{b}_{e}\, g^{ad}\, \nabla_{[a}\, R_{bc]d}{}^{e} = \tfrac{1}{3}\, [2\, \nabla_{a}\, G_{c}{}^{a} - 2\, \nabla_{b}\, (\nabla_{[a}\, Q_{c]}{}^{ab} + S_{ac}{}^{d}\, Q_{d}{}^{ab}) - \tag{91}$$
$$- 2\, Q_{[a}{}^{ab}\, R_{c]b} - Q_{b}{}^{ad}\, R_{cad}{}^{b}]$$
$$= 2\, S_{[ab}{}^{f}\, R_{c]f}{}^{ab}$$
$$= \tfrac{2}{3}\, (S_{ab}{}^{f}\, R_{cf}{}^{ab} + S_{bc}{}^{f}\, R_{af}{}^{ab} + S_{ca}{}^{f}\, R_{bf}{}^{ab})$$
$$= \tfrac{2}{3}\, [S_{ab}{}^{d}\, R_{cd}{}^{ab} - S_{bc}{}^{d}\, R_{ad}{}^{ba} +$$
$$+ 2\, S_{bc}{}^{d}\, (\nabla_{[a}\, Q_{d]}{}^{ab} + S_{ad}{}^{e}\, Q_{e}{}^{ab}) + S_{ca}{}^{f}\, R_{f}{}^{a}]$$
$$= \tfrac{2}{3}\, [S_{ab}{}^{d}\, R_{cd}{}^{ab} - 2\, S_{ac}{}^{b}\, R_{b}{}^{a} +$$
$$+ 2\, S_{bc}{}^{d}\, (\nabla_{[a}\, Q_{d]}{}^{ab} + S_{ad}{}^{e}\, Q_{e}{}^{ab})].$$

Another $3^{rd}$-order $1^{st}$ contraction of Eq. (89) is given by

$$\delta^{a}_{e}\, \nabla_{[a}\, R_{bc]d}{}^{e} = \nabla_{[a}\, R_{bc]d}{}^{a} \tag{92}$$
$$= \tfrac{1}{3}\, (\nabla_{a}\, R_{bcd}{}^{a} + \nabla_{b}\, R_{cad}{}^{a} + \nabla_{c}\, R_{abd}{}^{a})$$
$$= \tfrac{1}{3}\, (\nabla_{a}\, R_{bcd}{}^{a} - \nabla_{b}\, R_{cd} + \nabla_{c}\, R_{bd})$$
$$= \tfrac{1}{3}\, (\nabla_{a}\, R_{bcd}{}^{a} - 2\, \nabla_{[b}\, R_{c]d})$$
$$= 2\, S_{[ab}{}^{e}\, R_{c]ed}{}^{a}$$
$$= \tfrac{2}{3}\, (S_{ab}{}^{f}\, R_{cfd}{}^{a} + S_{bc}{}^{f}\, R_{afd}{}^{a} + S_{ca}{}^{f}\, R_{bfd}{}^{a})$$
$$= \tfrac{2}{3}\, (S_{ab}{}^{f}\, R_{cfd}{}^{a} + S_{bc}{}^{f}\, R_{fd} - S_{ac}{}^{f}\, R_{bfd}{}^{a})$$
$$= \tfrac{2}{3}\, (2\, S_{a[b}{}^{f}\, R_{c]fd}{}^{a} + S_{bc}{}^{f}\, R_{fd}),$$

one $1^{st}$-order $2^{nd}$ contraction of which is given by

$$g^{bd}\, \delta^{a}_{e}\, \nabla_{[a}\, R_{bc]d}{}^{e} = \tfrac{1}{3}\, [- 2\, \nabla_{a}\, G_{c}{}^{a} + 2\, \nabla_{a}\, (\nabla_{[b}\, Q_{c]}{}^{ba} + S_{bc}{}^{e}\, Q_{e}{}^{ba}) - \tag{93}$$
$$- Q_{a}{}^{be}\, R_{bce}{}^{a} + 2\, Q_{[b}{}^{be}\, R_{c]e}]$$
$$= 2\, S_{[ab}{}^{f}\, R_{c]f}{}^{ba}$$
$$= \tfrac{2}{3}\, [S_{ad}{}^{b}\, R_{cb}{}^{da} + 2\, S_{ac}{}^{b}\, R_{b}{}^{a} - 2\, S_{ac}{}^{b}\, (\nabla_{[d}\, Q_{b]}{}^{da} + S_{db}{}^{f}\, Q_{f}{}^{da})]$$

and another by

$$g^{cd}\, \delta^{a}_{e}\, \nabla_{[a}\, R_{bc]d}{}^{e} = - g^{cd}\, \delta^{a}_{e}\, \nabla_{[a}\, R_{cb]d}{}^{e} = - 2\, S_{[ab}{}^{f}\, R_{c]f}{}^{ba}, \tag{94}$$

which is simply the negative of Eq. (93).

Equation (91) or Eq. (93)—or Eq. (94) in view of Eq. (93)—yields the identity

$$\nabla_{b}\, G_{a}{}^{b} = \tfrac{1}{2}\, Q_{b}{}^{cd}\, R_{acd}{}^{b} + S_{bc}{}^{d}\, R_{ad}{}^{bc} + Q_{[b}{}^{bc}\, R_{a]c} - 2\, S_{ba}{}^{c}\, R_{c}{}^{b} + \tag{95}$$
$$+ \nabla_{c}\, (\nabla_{[b}\, Q_{a]}{}^{bc} + S_{ba}{}^{d}\, Q_{d}{}^{bc}) + 2\, S_{ca}{}^{d}\, (\nabla_{[b}\, Q_{d]}{}^{bc} + S_{bd}{}^{e}\, Q_{e}{}^{bc})$$

for the "mixed" components $G_{a}{}^{b}$ of the Einstein curvature tensor, which yields the identity

---

[59] Schouten, J. A., *ibid.*, p. 172; cf. p. 138.
[60] Schouten, J. A., *ibid.*, p. 146.

$$\nabla_b G^{ab} = \nabla_b g^{ac} G_c{}^b \tag{96}$$
$$= Q_b{}^{ac} G_c{}^b + g^{ac} \nabla_b G_c{}^b$$
$$= Q_b{}^{ac} G_c{}^b + \tfrac{1}{2} g^{ac} Q_b{}^{de} R_{cde}{}^b + g^{ac} S_{bd}{}^e R_{ce}{}^{bd} + g^{ac} Q_{[b}{}^{bd} R_{c]d} -$$
$$- 2 g^{ac} S_{bc}{}^d R_d{}^b + g^{ac} \nabla_d (\nabla_{[b} Q_{c]}{}^{bd} + S_{bc}{}^e Q_e{}^{bd}) +$$
$$+ 2 g^{ac} S_{dc}{}^e (\nabla_{[b} Q_{e]}{}^{bd} + S_{be}{}^f Q_f{}^{bd})$$
$$= Q_b{}^{ac} G_c{}^b + \tfrac{1}{2} Q_b{}^{de} R^a{}_{de}{}^b + S_{bd}{}^e R^a{}_e{}^{bd} + g^{ac} Q_{[b}{}^{bd} R_{c]d} - 2 S_b{}^{ad} R_d{}^b +$$
$$+ g^{ac} \nabla_d (\nabla_{[b} Q_{c]}{}^{bd} + S_{bc}{}^e Q_e{}^{bd}) + 2 S_d{}^{ae} (\nabla_{[b} Q_{e]}{}^{bd} + S_{be}{}^f Q_f{}^{bd})$$
$$= Q_b{}^{ac} G_c{}^b + (\tfrac{1}{2} Q_b{}^c{}_d - S_{bd}{}^c) R^a{}_c{}^{db} + \tfrac{1}{2} Q_b{}^{bd} R^a{}_d - \tfrac{1}{2} Q^{abd} R_{bd} -$$
$$- 2 S_b{}^{ac} R_c{}^b + g^{ab} \nabla_d (\nabla_{[c} Q_{b]}{}^{cd} + S_{cb}{}^e Q_e{}^{cd}) +$$
$$+ 2 S_b{}^{ac} (\nabla_{[d} Q_{c]}{}^{db} + S_{dc}{}^e Q_e{}^{db})$$
$$= Q_b{}^{ac} G_c{}^b + (\tfrac{1}{2} Q_b{}^c{}_d - S_{bd}{}^c) R^a{}_c{}^{db} + \tfrac{1}{2} Q^b{}_{bd} R^{ad} - \tfrac{1}{2} Q^a{}_{bd} R^{bd} -$$
$$- 2 S_b{}^{ac} R_c{}^b + g^{ab} \nabla_d (\nabla_{[c} Q_{b]}{}^{cd} + S_{cb}{}^e Q_e{}^{cd}) +$$
$$+ 2 S_b{}^{ac} (\nabla_{[d} Q_{c]}{}^{db} + S_{dc}{}^e Q_e{}^{db})$$
$$= Q_b{}^{ac} G_c{}^b + (\tfrac{1}{2} Q_b{}^c{}_d - S_{bd}{}^c) R^a{}_c{}^{db} + Q^{[b}{}_{bd} R^{a]d} - 2 S_b{}^{ac} R_c{}^b +$$
$$+ g^{ab} \nabla_d (\nabla_{[c} Q_{b]}{}^{cd} + S_{cb}{}^e Q_e{}^{cd}) + 2 S_b{}^{ac} (\nabla_{[d} Q_{c]}{}^{db} + S_{dc}{}^e Q_e{}^{db})$$

for the contravariant components $G^{ab}$ of the Einstein curvature tensor.

## APPENDIX C. CALCULATION OF GENERALIZED *n*-DIMENSIONAL GEODESIC DEVIATION EQUATIONS

Generalized $n$-dimensional geodesic deviation equations for the tangent and deviation vectors $u^a$ and $v^a$ are given (using anholonomic coordinates for a general linear connection) by

$$u^b \nabla_b u^c \nabla_c v^a = u^b \nabla_b (v^c \nabla_c u^a + 2 u^c v^d S_{cd}{}^a) \tag{97}$$
$$= u^b (\nabla_b v^c) \nabla_c u^a + u^b v^c \nabla_b \nabla_c u^a + 2 u^b \nabla_b u^c v^d S_{cd}{}^a$$
$$= u^b (\nabla_b v^c) \nabla_c u^a +$$
$$+ u^b v^c (\nabla_c \nabla_b u^a + u^b v^c R_{bcd}{}^a u^d - 2 S_{bc}{}^d \nabla_d u^a) +$$
$$+ 2 u^b u^c \nabla_b v^d S_{cd}{}^a$$
$$= v^b (\nabla_b u^c) \nabla_c u^a + 2 u^b v^c S_{bc}{}^d \nabla_d u^a - v^c (\nabla_c u^b) \nabla_b u^a +$$
$$+ u^b v^c R_{bcd}{}^a u^d - 2 u^b v^c S_{bc}{}^d \nabla_d u^a + 2 u^b u^c \nabla_b v^d S_{cd}{}^a$$
$$= u^b v^c R_{bcd}{}^a u^d + 2 u^b u^c S_{bd}{}^a \nabla_c v^d + 2 u^b u^c v^d \nabla_b S_{cd}{}^a$$
$$= u^b v^c u^d (R_{bcd}{}^a - 2 \nabla_b S_{cd}{}^a) + 2 u^b u^c S_{bd}{}^a \nabla_c v^d$$
$$= u^b v^c u^d (R_{bcd}{}^a - 2 \nabla_b S_{cd}{}^a) + 2 u^b v^c S_{bd}{}^a \nabla_c u^d +$$
$$+ 4 u^b v^c u^d S_{be}{}^a S_{dc}{}^e$$
$$= u^b v^c u^d (R_{bcd}{}^a - 2 \nabla_b S_{cd}{}^a - 4 S_{bc}{}^e S_{ed}{}^a) +$$
$$+ 2 u^b v^c S_{bd}{}^a \nabla_c u^d$$

using Ricci's identities[61] applied to the vector $u^a$, i.e.,

$$\nabla_{[b} \nabla_{c]} u^a = \tfrac{1}{2} R_{bcd}{}^a u^d - S_{bc}{}^d \nabla_d u^a, \tag{98}$$

and the formula

$$2 u^b v^c S_{bc}{}^a = u^b \nabla_b v^a - v^b \nabla_b u^a - (u^b \partial_b v^a - v^b \partial_b u^a), \tag{99}$$

which simplifies to

$$2 u^b v^c S_{bc}{}^a = u^b \nabla_b v^a - v^b \nabla_b u^a \tag{100}$$

since $u^a$ and $v^a$ commute, i.e., since

$$u^b \partial_b v^a = v^b \partial_b u^a. \tag{101}$$

## APPENDIX D. CALCULATION OF 5-DIMENSIONAL CONSERVATION AND GENERALIZED GEODESIC DEVIATION EQUATIONS

### 5-DIMENSIONAL CONSERVATION EQUATIONS

The two parts of the covariant divergence $\nabla_\beta T^{\alpha\beta}$ of the 5-dimensional energy-momentum tensor $T^{\alpha\beta}$ as defined (using anholonomic coordinates for a general linear connection) by

$$\nabla_\beta T^{\alpha\beta} = \partial_\beta T^{\alpha\beta} + \Gamma_\beta{}^\alpha{}_\gamma T^{\gamma\beta} + \Gamma_\beta{}^\beta{}_\gamma T^{\alpha\gamma} \tag{102}$$
$$= \tfrac{1}{k} \nabla_\beta G^{\alpha\beta}$$
$$= \tfrac{1}{k} \partial_\beta G^{\alpha\beta} + \tfrac{1}{k} \Gamma_\beta{}^\alpha{}_\gamma G^{\gamma\beta} + \tfrac{1}{k} \Gamma_\beta{}^\beta{}_\gamma G^{\alpha\gamma}$$
$$= \tfrac{1}{k} Q_\beta{}^{\alpha\gamma} G_\gamma{}^\beta + \tfrac{1}{k} (\tfrac{1}{2} Q_\beta{}^\gamma{}_\delta - S_{\beta\delta}{}^\gamma) R^\alpha{}_\gamma{}^{\delta\beta} + \tfrac{1}{k} Q^{[\beta}{}_{\beta\delta} R^{\alpha]\delta} -$$
$$- \tfrac{2}{k} S_\beta{}^{\alpha\gamma} R_\gamma{}^\beta + \tfrac{1}{k} g^{\alpha\beta} \nabla_\delta (\nabla_{[\gamma} Q_{\beta]}{}^{\gamma\delta} + S_{\gamma\beta}{}^\varepsilon Q_\varepsilon{}^{\gamma\delta}) +$$
$$+ \tfrac{2}{k} S_\beta{}^{\alpha\gamma} (\nabla_{[\delta} Q_{\gamma]}{}^{\delta\beta} + S_{\delta\gamma}{}^\varepsilon Q_\varepsilon{}^{\delta\beta})$$

are given, using Eq. (102) with $Q_\alpha{}^{\beta\gamma} = 0$, by

$$\nabla_\alpha T^{a\alpha} = \partial_\alpha T^{a\alpha} + \Gamma_\alpha{}^a{}_\beta T^{\beta\alpha} + \Gamma_\alpha{}^\alpha{}_\beta T^{a\beta} \tag{103}$$
$$= \partial_b T^{ab} + \partial_5 T^{a5} + \Gamma_b{}^a{}_c T^{cb} + \Gamma_b{}^a{}_5 T^{5b} + \Gamma_5{}^a{}_b T^{b5} +$$
$$+ \Gamma_5{}^a{}_5 T^{55} + \Gamma_b{}^b{}_c T^{ac} + \Gamma_b{}^b{}_5 T^{a5} + \Gamma_5{}^5{}_b T^{ab} + \Gamma_5{}^5{}_5 T^{a5}$$
$$= \tfrac{1}{k} \nabla_\alpha G^{a\alpha}$$
$$= \tfrac{1}{k} \partial_\alpha G^{a\alpha} + \tfrac{1}{k} \Gamma_\alpha{}^a{}_\beta G^{\beta\alpha} + \tfrac{1}{k} \Gamma_\alpha{}^\alpha{}_\beta G^{a\beta}$$
$$= \tfrac{1}{k} \partial_b G^{ab} + \tfrac{1}{k} \partial_5 G^{a5} + \tfrac{1}{k} \Gamma_b{}^a{}_c G^{cb} + \tfrac{1}{k} \Gamma_b{}^a{}_5 G^{5b} + \tfrac{1}{k} \Gamma_5{}^a{}_b G^{b5} +$$
$$+ \tfrac{1}{k} \Gamma_5{}^a{}_5 G^{55} + \tfrac{1}{k} \Gamma_b{}^b{}_c G^{ac} + \tfrac{1}{k} \Gamma_b{}^b{}_5 G^{a5} + \tfrac{1}{k} \Gamma_5{}^5{}_b G^{ab} +$$
$$+ \tfrac{1}{k} \Gamma_5{}^5{}_5 G^{a5}$$
$$= - \tfrac{1}{k} S_{\alpha\gamma}{}^\beta R^a{}_\beta{}^{\gamma\alpha} - \tfrac{2}{k} S_\alpha{}^{a\beta} R_\beta{}^\alpha$$
$$= - \tfrac{1}{k} S_{bc}{}^d R^a{}_d{}^{cb} - \tfrac{1}{k} S_{bc}{}^5 R^a{}_5{}^{cb} - \tfrac{1}{k} S_{b5}{}^c R^a{}_c{}^{5b} - \tfrac{1}{k} S_{b5}{}^5 R^a{}_5{}^{5b} -$$
$$- \tfrac{1}{k} S_{5b}{}^c R^a{}_c{}^{b5} - \tfrac{1}{k} S_{5b}{}^5 R^a{}_5{}^{b5} - \tfrac{1}{k} S_{55}{}^b R^a{}_b{}^{55} -$$
$$- \tfrac{1}{k} S_{55}{}^5 R^a{}_5{}^{55} - \tfrac{2}{k} S_b{}^{ac} R_c{}^b - \tfrac{2}{k} S_b{}^{a5} R_5{}^b - \tfrac{2}{k} S_5{}^{ab} R_b{}^5 -$$
$$- \tfrac{2}{k} S_5{}^{a5} R_5{}^5$$
$$= - \tfrac{1}{k} S_{bc}{}^d R^a{}_d{}^{cb} - \tfrac{1}{k} S_{bc}{}^5 R^a{}_5{}^{cb} - \tfrac{1}{k} S_{b5}{}^c R^a{}_c{}^{5b} - \tfrac{1}{k} S_{b5}{}^5 R^a{}_5{}^{5b} -$$
$$- \tfrac{1}{k} S_{5b}{}^c R^a{}_c{}^{b5} - \tfrac{1}{k} S_{5b}{}^5 R^a{}_5{}^{b5} - \tfrac{2}{k} S_b{}^{ac} R_c{}^b - \tfrac{2}{k} S_b{}^{a5} R_5{}^b -$$
$$- \tfrac{2}{k} S_5{}^{ab} R_b{}^5 - \tfrac{2}{k} S_5{}^{a5} R_5{}^5$$

and

$$\nabla_\alpha T^{5\alpha} = \partial_\alpha T^{5\alpha} + \Gamma_\alpha{}^5{}_\beta T^{\beta\alpha} + \Gamma_\alpha{}^\alpha{}_\beta T^{5\beta} \tag{104}$$
$$= \partial_a T^{5a} + \partial_5 T^{55} + \Gamma_a{}^5{}_b T^{ba} + \Gamma_a{}^5{}_5 T^{5a} + \Gamma_5{}^5{}_a T^{a5} +$$
$$+ \Gamma_5{}^5{}_5 T^{55} + \Gamma_a{}^a{}_b T^{5b} + \Gamma_a{}^a{}_5 T^{55} + \Gamma_5{}^5{}_a T^{5a} + \Gamma_5{}^5{}_5 T^{55}$$
$$= \tfrac{1}{k} \nabla_\alpha G^{5\alpha}$$
$$= \tfrac{1}{k} \partial_\alpha G^{5\alpha} + \tfrac{1}{k} \Gamma_\alpha{}^5{}_\beta G^{\beta\alpha} + \tfrac{1}{k} \Gamma_\alpha{}^\alpha{}_\beta G^{5\beta}$$
$$= \tfrac{1}{k} \partial_a G^{5a} + \tfrac{1}{k} \partial_5 G^{55} + \tfrac{1}{k} \Gamma_a{}^5{}_c G^{ca} + \tfrac{1}{k} \Gamma_a{}^5{}_5 G^{5a} + \tfrac{1}{k} \Gamma_5{}^5{}_a G^{a5} +$$
$$+ \tfrac{1}{k} \Gamma_5{}^5{}_5 G^{55} + \tfrac{1}{k} \Gamma_a{}^a{}_c G^{5c} + \tfrac{1}{k} \Gamma_a{}^a{}_5 G^{55} + \tfrac{1}{k} \Gamma_5{}^5{}_a G^{5a} +$$
$$+ \tfrac{1}{k} \Gamma_5{}^5{}_5 G^{55}$$
$$= - \tfrac{1}{k} S_{\alpha\gamma}{}^\beta R^5{}_\beta{}^{\gamma\alpha} - \tfrac{2}{k} S_\alpha{}^{5\beta} R_\beta{}^\alpha$$

[61] Schouten, J. A., *ibid.*, p. 139.

$$\begin{aligned}
&= -\tfrac{1}{k} S_{ab}{}^{c} R^{5}{}_{c}{}^{ba} - \tfrac{1}{k} S_{ab}{}^{5} R^{5}{}_{5}{}^{ba} - \tfrac{1}{k} S_{a5}{}^{b} R^{5}{}_{b}{}^{5a} - \tfrac{1}{k} S_{a5}{}^{5} R^{5}{}_{5}{}^{5a} - \\
&\quad - \tfrac{1}{k} S_{5a}{}^{b} R^{5}{}_{b}{}^{a5} - \tfrac{1}{k} S_{5a}{}^{5} R^{5}{}_{5}{}^{a5} - \tfrac{1}{k} S_{55}{}^{a} R^{5}{}_{a}{}^{55} - \\
&\quad - \tfrac{1}{k} S_{55}{}^{5} R^{5}{}_{5}{}^{55} - \tfrac{2}{k} S_{a}{}^{5b} R_{b}{}^{a} - \tfrac{2}{k} S_{a}{}^{55} R_{5}{}^{a} - \tfrac{2}{k} S_{5}{}^{5a} R_{a}{}^{5} - \\
&\quad - \tfrac{2}{k} S_{5}{}^{55} R_{5}{}^{5} \\
&= -\tfrac{1}{k} S_{ab}{}^{c} R^{5}{}_{c}{}^{ba} - \tfrac{1}{k} S_{ab}{}^{5} R^{5}{}_{5}{}^{ba} - \tfrac{1}{k} S_{a5}{}^{b} R^{5}{}_{b}{}^{5a} - \tfrac{1}{k} S_{a5}{}^{5} R^{5}{}_{5}{}^{5a} - \\
&\quad - \tfrac{1}{k} S_{5a}{}^{b} R^{5}{}_{b}{}^{a5} - \tfrac{1}{k} S_{5a}{}^{5} R^{5}{}_{5}{}^{a5} - \tfrac{2}{k} S_{a}{}^{5b} R_{b}{}^{a} - \tfrac{2}{k} S_{a}{}^{55} R_{5}{}^{a} - \\
&\quad - \tfrac{2}{k} S_{5}{}^{5a} R_{a}{}^{5} - \tfrac{2}{k} S_{5}{}^{55} R_{5}{}^{5},
\end{aligned}$$

respectively.

### 5-DIMENSIONAL GEODESIC DEVIATION EQUATIONS

The two parts of the 5-dimensional geodesic deviation equations as defined (using anholonomic coordinates for a general linear connection) by

$$\begin{aligned}
u^{\beta} \nabla_{\beta} u^{\gamma} \nabla_{\gamma} v^{\alpha} &= u^{\beta} v^{\gamma} u^{\delta} (R_{\beta\gamma\delta}{}^{\alpha} - 2\,\nabla_{\beta} S_{\gamma\delta}{}^{\alpha}) + 2\,u^{\beta} u^{\gamma} S_{\beta\delta}{}^{\alpha} \nabla_{\gamma} v^{\delta} \qquad (105) \\
&= u^{\beta} v^{\gamma} u^{\delta} (R_{\beta\gamma\delta}{}^{\alpha} - 2\,\nabla_{\beta} S_{\gamma\delta}{}^{\alpha} - 4\,S_{\beta\varepsilon}{}^{\alpha} S_{\gamma\delta}{}^{\varepsilon}) + \\
&\quad + 2\,u^{\beta} v^{\gamma} S_{\beta\delta}{}^{\alpha} \nabla_{\gamma} u^{\delta}
\end{aligned}$$

are given by

$$u^{\alpha} \nabla_{\alpha} u^{\beta} \nabla_{\beta} v^{a} = \qquad (106)$$

$$\begin{aligned}
&= u^{b} \nabla_{b} u^{c} \nabla_{c} v^{a} + u^{b} \nabla_{b} u^{5} \nabla_{5} v^{a} + u^{5} \nabla_{5} u^{b} \nabla_{b} v^{a} + u^{5} \nabla_{5} u^{5} \nabla_{5} v^{a} \\
&= u^{\alpha} v^{\beta} u^{\gamma} (R_{\alpha\beta\gamma}{}^{a} - 2\,\nabla_{\alpha} S_{\beta\gamma}{}^{a}) + 2\,u^{\alpha} u^{\beta} S_{\alpha\gamma}{}^{a} \nabla_{\beta} v^{\gamma} \\
&= u^{b} v^{c} u^{d} (R_{bcd}{}^{a} - 2\,\nabla_{b} S_{cd}{}^{a}) + u^{b} v^{c} u^{5} (R_{bc5}{}^{a} - 2\,\nabla_{b} S_{c5}{}^{a}) + \\
&\quad + u^{b} v^{5} u^{c} (R_{b5c}{}^{a} - 2\,\nabla_{b} S_{5c}{}^{a}) + u^{b} v^{5} u^{5} (R_{b55}{}^{a} - 2\,\nabla_{b} S_{55}{}^{a}) + \\
&\quad + u^{5} v^{b} u^{c} (R_{5bc}{}^{a} - 2\,\nabla_{5} S_{bc}{}^{a}) + u^{5} v^{b} u^{5} (R_{5b5}{}^{a} - 2\,\nabla_{5} S_{b5}{}^{a}) + \\
&\quad + u^{5} v^{5} u^{b} (R_{55b}{}^{a} - 2\,\nabla_{5} S_{5b}{}^{a}) + u^{5} v^{5} u^{5} (R_{555}{}^{a} - 2\,\nabla_{5} S_{55}{}^{a}) + \\
&\quad + 2\,u^{b} u^{c} S_{bd}{}^{a} \nabla_{c} v^{d} + 2\,u^{b} u^{c} S_{b5}{}^{a} \nabla_{c} v^{5} + 2\,u^{b} u^{5} S_{bc}{}^{a} \nabla_{5} v^{c} + \\
&\quad + 2\,u^{b} u^{5} S_{b5}{}^{a} \nabla_{5} v^{5} + 2\,u^{5} u^{b} S_{5c}{}^{a} \nabla_{b} v^{c} + 2\,u^{5} u^{b} S_{55}{}^{a} \nabla_{b} v^{5} + \\
&\quad + 2\,u^{5} u^{5} S_{5b}{}^{a} \nabla_{5} v^{b} + 2\,u^{5} u^{5} S_{55}{}^{a} \nabla_{5} v^{5} \\
&= u^{b} v^{c} u^{d} (R_{bcd}{}^{a} - 2\,\nabla_{b} S_{cd}{}^{a}) + u^{b} v^{c} u^{5} (R_{bc5}{}^{a} - 2\,\nabla_{b} S_{c5}{}^{a}) + \\
&\quad + u^{b} v^{5} u^{c} (R_{b5c}{}^{a} - 2\,\nabla_{b} S_{5c}{}^{a}) + u^{b} v^{5} u^{5} R_{b55}{}^{a} + \\
&\quad + u^{5} v^{b} u^{c} (R_{5bc}{}^{a} - 2\,\nabla_{5} S_{bc}{}^{a}) + u^{5} v^{b} u^{5} (R_{5b5}{}^{a} - 2\,\nabla_{5} S_{b5}{}^{a}) - \\
&\quad - 2\,u^{5} v^{5} u^{b} \nabla_{5} S_{5b}{}^{a} + 2\,u^{b} u^{c} S_{bd}{}^{a} \nabla_{c} v^{d} + 2\,u^{b} u^{c} S_{b5}{}^{a} \nabla_{c} v^{5} + \\
&\quad + 2\,u^{b} u^{5} S_{bc}{}^{a} \nabla_{5} v^{c} + 2\,u^{b} u^{5} S_{b5}{}^{a} \nabla_{5} v^{5} + 2\,u^{5} u^{b} S_{5c}{}^{a} \nabla_{b} v^{c} + \\
&\quad + 2\,u^{5} u^{5} S_{5b}{}^{a} \nabla_{5} v^{b} \\
&= u^{\alpha} v^{\beta} u^{\gamma} (R_{\alpha\beta\gamma}{}^{a} - 2\,\nabla_{\alpha} S_{\beta\gamma}{}^{a} - 4\,S_{\alpha\delta}{}^{a} S_{\beta\gamma}{}^{\delta}) + 2\,u^{\alpha} v^{\beta} S_{\alpha\gamma}{}^{a} \nabla_{\beta} u^{\gamma} \\
&= u^{b} v^{c} u^{d} (R_{bcd}{}^{a} - 2\,\nabla_{b} S_{cd}{}^{a}) + u^{b} v^{c} u^{5} (R_{bc5}{}^{a} - 2\,\nabla_{b} S_{c5}{}^{a}) + \\
&\quad + u^{b} v^{5} u^{c} (R_{b5c}{}^{a} - 2\,\nabla_{b} S_{5c}{}^{a}) + u^{b} v^{5} u^{5} (R_{b55}{}^{a} - 2\,\nabla_{b} S_{55}{}^{a}) + \\
&\quad + u^{5} v^{b} u^{c} (R_{5bc}{}^{a} - 2\,\nabla_{5} S_{bc}{}^{a}) + u^{5} v^{b} u^{5} (R_{5b5}{}^{a} - 2\,\nabla_{5} S_{b5}{}^{a}) + \\
&\quad + u^{5} v^{5} u^{b} (R_{55b}{}^{a} - 2\,\nabla_{5} S_{5b}{}^{a}) + u^{5} v^{5} u^{5} (R_{555}{}^{a} - 2\,\nabla_{5} S_{55}{}^{a}) - \\
&\quad - 4\,u^{b} v^{c} u^{d} S_{be}{}^{a} S_{cd}{}^{e} - 4\,u^{b} v^{c} u^{d} S_{b5}{}^{a} S_{cd}{}^{5} - 4\,u^{b} v^{c} u^{5} S_{bd}{}^{a} S_{c5}{}^{d} - \\
&\quad - 4\,u^{b} v^{c} u^{5} S_{b5}{}^{a} S_{c5}{}^{5} - 4\,u^{b} v^{5} u^{c} S_{bd}{}^{a} S_{5c}{}^{d} - 4\,u^{b} v^{5} u^{c} S_{b5}{}^{a} S_{5c}{}^{5} - \\
&\quad - 4\,u^{b} v^{5} u^{5} S_{bc}{}^{a} S_{55}{}^{c} - 4\,u^{b} v^{5} u^{5} S_{b5}{}^{a} S_{55}{}^{5} - 4\,u^{5} v^{b} u^{c} S_{5d}{}^{a} S_{bc}{}^{d} - \\
&\quad - 4\,u^{5} v^{b} u^{c} S_{55}{}^{a} S_{bc}{}^{5} - 4\,u^{5} v^{b} u^{5} S_{5c}{}^{a} S_{b5}{}^{c} - 4\,u^{5} v^{b} u^{5} S_{55}{}^{a} S_{b5}{}^{5} - \\
&\quad - 4\,u^{5} v^{5} u^{b} S_{5c}{}^{a} S_{5b}{}^{c} - 4\,u^{5} v^{5} u^{b} S_{55}{}^{a} S_{5b}{}^{5} - 4\,u^{5} v^{5} u^{5} S_{5b}{}^{a} S_{55}{}^{b} - \\
&\quad - 4\,u^{5} v^{5} u^{5} S_{55}{}^{a} S_{55}{}^{5} + 2\,u^{b} v^{c} S_{bd}{}^{a} \nabla_{c} u^{d} + 2\,u^{b} v^{c} S_{b5}{}^{a} \nabla_{c} u^{5} + \\
&\quad + 2\,u^{b} v^{5} S_{bc}{}^{a} \nabla_{5} u^{c} + 2\,u^{b} v^{5} S_{b5}{}^{a} \nabla_{5} u^{5} + 2\,u^{5} v^{b} S_{5c}{}^{a} \nabla_{b} u^{c} + \\
&\quad + 2\,u^{5} v^{b} S_{55}{}^{a} \nabla_{b} u^{5} + 2\,u^{5} v^{5} S_{5b}{}^{a} \nabla_{5} u^{b} + 2\,u^{5} v^{5} S_{55}{}^{a} \nabla_{5} u^{5} \\
&= u^{b} v^{c} u^{d} (R_{bcd}{}^{a} - 2\,\nabla_{b} S_{cd}{}^{a}) + u^{b} v^{c} u^{5} (R_{bc5}{}^{a} - 2\,\nabla_{b} S_{c5}{}^{a}) + \\
&\quad + u^{b} v^{5} u^{c} (R_{b5c}{}^{a} - 2\,\nabla_{b} S_{5c}{}^{a}) + u^{b} v^{5} u^{5} R_{b55}{}^{a} + \\
&\quad + u^{5} v^{b} u^{c} (R_{5bc}{}^{a} - 2\,\nabla_{5} S_{bc}{}^{a}) + u^{5} v^{b} u^{5} (R_{5b5}{}^{a} - 2\,\nabla_{5} S_{b5}{}^{a}) - \\
&\quad - 2\,u^{5} v^{5} u^{b} \nabla_{5} S_{5b}{}^{a} - 4\,u^{b} v^{c} u^{d} S_{be}{}^{a} S_{cd}{}^{e} - 4\,u^{b} v^{c} u^{d} S_{b5}{}^{a} S_{cd}{}^{5} - \\
&\quad - 4\,u^{b} v^{c} u^{5} S_{bd}{}^{a} S_{c5}{}^{d} - 4\,u^{b} v^{c} u^{5} S_{b5}{}^{a} S_{c5}{}^{5} - 4\,u^{b} v^{5} u^{c} S_{bd}{}^{a} S_{5c}{}^{d} - \\
&\quad - 4\,u^{b} v^{5} u^{c} S_{b5}{}^{a} S_{5c}{}^{5} - 4\,u^{b} v^{5} u^{5} S_{bc}{}^{a} S_{55}{}^{c} - 4\,u^{b} v^{5} u^{5} S_{b5}{}^{a} S_{55}{}^{5} - \\
&\quad - 4\,u^{5} v^{b} u^{c} S_{5d}{}^{a} S_{bc}{}^{d} - 4\,u^{5} v^{b} u^{5} S_{5c}{}^{a} S_{b5}{}^{c} - 4\,u^{5} v^{5} u^{b} S_{5c}{}^{a} S_{5b}{}^{c} + \\
&\quad + 2\,u^{b} v^{c} S_{bd}{}^{a} \nabla_{c} u^{d} + 2\,u^{b} v^{c} S_{b5}{}^{a} \nabla_{c} u^{5} + 2\,u^{b} v^{5} S_{bc}{}^{a} \nabla_{5} u^{c} + \\
&\quad + 2\,u^{b} v^{5} S_{b5}{}^{a} \nabla_{5} u^{5} + 2\,u^{5} v^{b} S_{5c}{}^{a} \nabla_{b} u^{c} + 2\,u^{5} v^{5} S_{5b}{}^{a} \nabla_{5} u^{b}
\end{aligned}$$

and

$$u^{\alpha} \nabla_{\alpha} u^{\beta} \nabla_{\beta} v^{5} = \qquad (107)$$

$$\begin{aligned}
&= u^{a} \nabla_{a} u^{b} \nabla_{b} v^{5} + u^{a} \nabla_{a} u^{5} \nabla_{5} v^{5} + u^{5} \nabla_{5} u^{a} \nabla_{a} v^{5} + u^{5} \nabla_{5} u^{5} \nabla_{5} v^{5} \\
&= u^{\alpha} v^{\beta} u^{\gamma} (R_{\alpha\beta\gamma}{}^{5} - 2\,\nabla_{\alpha} S_{\beta\gamma}{}^{5}) + 2\,u^{\alpha} u^{\beta} S_{\alpha\gamma}{}^{5} \nabla_{\beta} v^{\gamma} \\
&= u^{a} v^{b} u^{c} (R_{abc}{}^{5} - 2\,\nabla_{a} S_{bc}{}^{5}) + u^{a} v^{b} u^{5} (R_{ab5}{}^{5} - 2\,\nabla_{a} S_{b5}{}^{5}) + \\
&\quad + u^{a} v^{5} u^{b} (R_{a5b}{}^{5} - 2\,\nabla_{a} S_{5b}{}^{5}) + u^{a} v^{5} u^{5} (R_{a55}{}^{5} - 2\,\nabla_{a} S_{55}{}^{5}) + \\
&\quad + u^{5} v^{a} u^{b} (R_{5ab}{}^{5} - 2\,\nabla_{5} S_{ab}{}^{5}) + u^{5} v^{a} u^{5} (R_{5a5}{}^{5} - 2\,\nabla_{5} S_{a5}{}^{5}) + \\
&\quad + u^{5} v^{5} u^{a} (R_{55a}{}^{5} - 2\,\nabla_{5} S_{5a}{}^{5}) + u^{5} v^{5} u^{5} (R_{555}{}^{5} - 2\,\nabla_{5} S_{55}{}^{5}) + \\
&\quad + 2\,u^{a} u^{b} S_{ac}{}^{5} \nabla_{b} v^{c} + 2\,u^{a} u^{b} S_{a5}{}^{5} \nabla_{b} v^{5} + 2\,u^{a} u^{5} S_{ab}{}^{5} \nabla_{5} v^{b} + \\
&\quad + 2\,u^{a} u^{5} S_{a5}{}^{5} \nabla_{5} v^{5} + 2\,u^{5} u^{a} S_{5b}{}^{5} \nabla_{a} v^{b} + 2\,u^{5} u^{a} S_{55}{}^{5} \nabla_{a} v^{5} + \\
&\quad + 2\,u^{5} u^{5} S_{5a}{}^{5} \nabla_{5} v^{a} + 2\,u^{5} u^{5} S_{55}{}^{5} \nabla_{5} v^{5} \\
&= u^{a} v^{b} u^{c} (R_{abc}{}^{5} - 2\,\nabla_{a} S_{bc}{}^{5}) + u^{a} v^{b} u^{5} (R_{ab5}{}^{5} - 2\,\nabla_{a} S_{b5}{}^{5}) + \\
&\quad + u^{a} v^{5} u^{b} (R_{a5b}{}^{5} - 2\,\nabla_{a} S_{5b}{}^{5}) + u^{a} v^{5} u^{5} R_{a55}{}^{5} + \\
&\quad + u^{5} v^{a} u^{b} (R_{5ab}{}^{5} - 2\,\nabla_{5} S_{ab}{}^{5}) + u^{5} v^{a} u^{5} (R_{5a5}{}^{5} - 2\,\nabla_{5} S_{a5}{}^{5}) - \\
&\quad - 2\,u^{5} v^{5} u^{a} \nabla_{5} S_{5a}{}^{5} + 2\,u^{a} u^{b} S_{ac}{}^{5} \nabla_{b} v^{c} + 2\,u^{a} u^{b} S_{a5}{}^{5} \nabla_{b} v^{5} + \\
&\quad + 2\,u^{a} u^{5} S_{ab}{}^{5} \nabla_{5} v^{b} + 2\,u^{a} u^{5} S_{a5}{}^{5} \nabla_{5} v^{5} + 2\,u^{5} u^{a} S_{5b}{}^{5} \nabla_{a} v^{b} + \\
&\quad + 2\,u^{5} u^{5} S_{5a}{}^{5} \nabla_{5} v^{a} \\
&= u^{\alpha} v^{\beta} u^{\gamma} (R_{\alpha\beta\gamma}{}^{5} - 2\,\nabla_{\alpha} S_{\beta\gamma}{}^{5} - 4\,S_{\alpha\delta}{}^{5} S_{\beta\gamma}{}^{\delta}) + 2\,u^{\alpha} v^{\beta} S_{\alpha\gamma}{}^{5} \nabla_{\beta} u^{\gamma} \\
&= u^{a} v^{b} u^{c} (R_{abc}{}^{5} - 2\,\nabla_{a} S_{bc}{}^{5}) + u^{a} v^{b} u^{5} (R_{ab5}{}^{5} - 2\,\nabla_{a} S_{b5}{}^{5}) + \\
&\quad + u^{a} v^{5} u^{b} (R_{a5b}{}^{5} - 2\,\nabla_{a} S_{5b}{}^{5}) + u^{a} v^{5} u^{5} (R_{a55}{}^{5} - 2\,\nabla_{a} S_{55}{}^{5}) + \\
&\quad + u^{5} v^{a} u^{b} (R_{5ab}{}^{5} - 2\,\nabla_{5} S_{ab}{}^{5}) + u^{5} v^{a} u^{5} (R_{5a5}{}^{5} - 2\,\nabla_{5} S_{a5}{}^{5}) + \\
&\quad + u^{5} v^{5} u^{a} (R_{55a}{}^{5} - 2\,\nabla_{5} S_{5a}{}^{5}) + u^{5} v^{5} u^{5} (R_{555}{}^{5} - 2\,\nabla_{5} S_{55}{}^{5}) - \\
&\quad - 4\,u^{a} v^{b} u^{c} S_{ae}{}^{5} S_{bc}{}^{e} - 4\,u^{a} v^{b} u^{c} S_{a5}{}^{5} S_{bc}{}^{5} - 4\,u^{a} v^{b} u^{5} S_{ac}{}^{5} S_{b5}{}^{c} - \\
&\quad - 4\,u^{a} v^{b} u^{5} S_{a5}{}^{5} S_{b5}{}^{5} - 4\,u^{a} v^{5} u^{b} S_{ac}{}^{5} S_{5b}{}^{c} - 4\,u^{a} v^{5} u^{b} S_{a5}{}^{5} S_{5b}{}^{5} - \\
&\quad - 4\,u^{a} v^{5} u^{5} S_{ab}{}^{5} S_{55}{}^{b} - 4\,u^{a} v^{5} u^{5} S_{a5}{}^{5} S_{55}{}^{5} - 4\,u^{5} v^{a} u^{b} S_{5c}{}^{5} S_{ab}{}^{c} - \\
&\quad - 4\,u^{5} v^{a} u^{b} S_{55}{}^{5} S_{ab}{}^{5} - 4\,u^{5} v^{a} u^{5} S_{5b}{}^{5} S_{a5}{}^{b} - 4\,u^{5} v^{a} u^{5} S_{55}{}^{5} S_{a5}{}^{5} - \\
&\quad - 4\,u^{5} v^{5} u^{a} S_{5b}{}^{5} S_{5a}{}^{b} - 4\,u^{5} v^{5} u^{a} S_{55}{}^{5} S_{5a}{}^{5} - 4\,u^{5} v^{5} u^{5} S_{5a}{}^{5} S_{55}{}^{a} - \\
&\quad - 4\,u^{5} v^{5} u^{5} S_{55}{}^{5} S_{55}{}^{5} + 2\,u^{a} v^{b} S_{ac}{}^{5} \nabla_{b} u^{c} + 2\,u^{a} v^{b} S_{a5}{}^{5} \nabla_{b} u^{5} + \\
&\quad + 2\,u^{a} v^{5} S_{ab}{}^{5} \nabla_{5} u^{b} + 2\,u^{a} v^{5} S_{a5}{}^{5} \nabla_{5} u^{5} + 2\,u^{5} v^{a} S_{5b}{}^{5} \nabla_{a} u^{b} + \\
&\quad + 2\,u^{5} v^{a} S_{55}{}^{5} \nabla_{a} u^{5} + 2\,u^{5} v^{5} S_{5a}{}^{5} \nabla_{5} u^{a} + 2\,u^{5} v^{5} S_{55}{}^{5} \nabla_{5} u^{5}
\end{aligned}$$

$$
\begin{aligned}
&= u^a v^b u^c (R_{abc}{}^5 - 2\, \nabla_a S_{bc}{}^5) + u^a v^b u^5 (R_{ab5}{}^5 - 2\, \nabla_a S_{b5}{}^5) + \\
&\quad + u^a v^5 u^b (R_{a5b}{}^5 - 2\, \nabla_a S_{5b}{}^5) + u^a v^5 u^5 R_{a55}{}^5 + \\
&\quad + u^5 v^a u^b (R_{5ab}{}^5 - 2\, \nabla_5 S_{ab}{}^5) + u^5 v^a u^5 (R_{5a5}{}^5 - 2\, \nabla_5 S_{a5}{}^5) - \\
&\quad - 2\, u^5 v^5 u^a \nabla_5 S_{5a}{}^5 - 4\, u^a v^b u^c S_{ae}{}^5 S_{bc}{}^e - 4\, u^a v^b u^c S_{a5}{}^5 S_{bc}{}^5 - \\
&\quad - 4\, u^a v^b u^5 S_{ac}{}^5 S_{b5}{}^c - 4\, u^a v^b u^5 S_{a5}{}^5 S_{b5}{}^5 - 4\, u^a v^5 u^b S_{ac}{}^5 S_{5b}{}^c - \\
&\quad - 4\, u^a v^5 u^b S_{a5}{}^5 S_{5b}{}^5 - 4\, u^a v^5 u^5 S_{ab}{}^5 S_{55}{}^b - 4\, u^a v^5 u^5 S_{a5}{}^5 S_{55}{}^5 - \\
&\quad - 4\, u^5 v^a u^b S_{5c}{}^5 S_{ab}{}^c - 4\, u^5 v^a u^5 S_{5b}{}^5 S_{a5}{}^b - 4\, u^5 v^5 u^a S_{5b}{}^5 S_{5a}{}^b + \\
&\quad + 2\, u^a v^b S_{ac}{}^5 \nabla_b u^c + 2\, u^a v^b S_{a5}{}^5 \nabla_b u^5 + 2\, u^a v^5 S_{ab}{}^5 \nabla_5 u^b + \\
&\quad + 2\, u^a v^5 S_{a5}{}^5 \nabla_5 u^5 + 2\, u^5 v^a S_{5b}{}^5 \nabla_a u^b + 2\, u^5 v^5 S_{5a}{}^5 \nabla_5 u^a,
\end{aligned}
$$

respectively.

## Appendix E. Expressions for 5-Dimensional Force Density Vectors

### 5-Dimensional Ponderomotive Force Density Vector

The two parts of the 5-dimensional ponderomotive force density vector as defined (using anholonomic coordinates for a general linear connection) by

$$
\begin{aligned}
F_{(1)}{}^\alpha &= \nabla_\beta T^{\alpha\beta} \qquad (108) \\
&= \partial_\beta T^{\alpha\beta} + \Gamma_\beta{}^\alpha{}_\gamma T^{\gamma\beta} + \Gamma_\beta{}^\beta{}_\gamma T^{\alpha\gamma} \\
&= \tfrac{1}{k} \nabla_\beta G^{\alpha\beta} \\
&= \tfrac{1}{k} \partial_\beta G^{\alpha\beta} + \tfrac{1}{k} \Gamma_\beta{}^\alpha{}_\gamma G^{\gamma\beta} + \tfrac{1}{k} \Gamma_\beta{}^\beta{}_\gamma G^{\alpha\gamma} \\
&= \tfrac{1}{k} Q_\beta{}^{\alpha\gamma} G_\gamma{}^\beta + \tfrac{1}{k} (\tfrac{1}{2} Q_\beta{}^\gamma{}_\delta - S_{\beta\delta}{}^\gamma) R^\alpha{}_\gamma{}^{\delta\beta} + \tfrac{1}{k} Q^{[\beta}{}_{\beta\delta} R^{\alpha]\delta} - \\
&\quad - \tfrac{2}{k} S_\beta{}^{\alpha\gamma} R_\gamma{}^\beta + \tfrac{1}{k} g^{\alpha\beta} \nabla_\delta (\nabla_{[\gamma} Q_{\beta]}{}^{\gamma\delta} + S_{\gamma\beta}{}^\varepsilon Q_\varepsilon{}^{\gamma\delta}) + \\
&\quad + \tfrac{2}{k} S_\beta{}^{\alpha\gamma} (\nabla_{[\delta} Q_{\gamma]}{}^{\delta\beta} + S_{\delta\gamma}{}^\varepsilon Q_\varepsilon{}^{\delta\beta})
\end{aligned}
$$

are given, using Eq. (108) with $Q_\alpha{}^{\beta\gamma} = 0$, by

$$
\begin{aligned}
F_{(1)}{}^a &= \nabla_\alpha T^{a\alpha} \qquad (109) \\
&= \partial_\alpha T^{a\alpha} + \Gamma_\alpha{}^a{}_\beta T^{\beta\alpha} + \Gamma_\alpha{}^\alpha{}_\beta T^{a\beta} \\
&= \partial_b T^{ab} + \partial_5 T^{a5} + \Gamma_b{}^a{}_c T^{cb} + \Gamma_b{}^a{}_5 T^{5b} + \Gamma_5{}^a{}_b T^{b5} + \Gamma_5{}^a{}_5 T^{55} + \\
&\quad + \Gamma_b{}^b{}_c T^{ac} + \Gamma_b{}^b{}_5 T^{a5} + \Gamma_5{}^5{}_b T^{ab} + \Gamma_5{}^5{}_5 T^{a5} \\
&= \tfrac{1}{k} \nabla_\alpha G^{a\alpha} \\
&= \tfrac{1}{k} \partial_\alpha G^{a\alpha} + \tfrac{1}{k} \Gamma_\alpha{}^a{}_\beta G^{\beta\alpha} + \tfrac{1}{k} \Gamma_\alpha{}^\alpha{}_\beta G^{a\beta} \\
&= \tfrac{1}{k} \partial_b G^{ab} + \tfrac{1}{k} \partial_5 G^{a5} + \tfrac{1}{k} \Gamma_b{}^a{}_c G^{cb} + \tfrac{1}{k} \Gamma_b{}^a{}_5 G^{5b} + \tfrac{1}{k} \Gamma_5{}^a{}_b G^{b5} + \\
&\quad + \tfrac{1}{k} \Gamma_5{}^a{}_5 G^{55} + \tfrac{1}{k} \Gamma_b{}^b{}_c G^{ac} + \tfrac{1}{k} \Gamma_b{}^b{}_5 G^{a5} + \tfrac{1}{k} \Gamma_5{}^5{}_b G^{ab} + \\
&\quad + \tfrac{1}{k} \Gamma_5{}^5{}_5 G^{a5} \\
&= -\tfrac{1}{k} S_{bc}{}^d R^a{}_d{}^{cb} - \tfrac{1}{k} S_{bc}{}^5 R^a{}_5{}^{cb} - \tfrac{1}{k} S_{b5}{}^c R^a{}_c{}^{5b} - \tfrac{1}{k} S_{b5}{}^5 R^a{}_5{}^{5b} - \\
&\quad - \tfrac{1}{k} S_{5b}{}^c R^a{}_c{}^{b5} - \tfrac{1}{k} S_{5b}{}^5 R^a{}_5{}^{b5} - \tfrac{2}{k} S_b{}^{ac} R_c{}^b - \tfrac{2}{k} S_b{}^{a5} R_5{}^b - \\
&\quad - \tfrac{2}{k} S_5{}^{ab} R_b{}^5 - \tfrac{2}{k} S_5{}^{a5} R_5{}^5
\end{aligned}
$$

and

$$
\begin{aligned}
F_{(1)}{}^5 &= \nabla_\alpha T^{5\alpha} \qquad (110) \\
&= \partial_\alpha T^{5\alpha} + \Gamma_\alpha{}^5{}_\beta T^{\beta\alpha} + \Gamma_\alpha{}^\alpha{}_\beta T^{5\beta} \\
&= \partial_a T^{5a} + \partial_5 T^{55} + \Gamma_a{}^5{}_b T^{ba} + \Gamma_a{}^5{}_5 T^{5a} + \Gamma_5{}^5{}_a T^{a5} + \\
&\quad + \Gamma_5{}^5{}_5 T^{55} + \Gamma_a{}^a{}_b T^{5b} + \Gamma_a{}^a{}_5 T^{55} + \Gamma_5{}^5{}_a T^{5a} + \Gamma_5{}^5{}_5 T^{55} \\
&= \tfrac{1}{k} \nabla_\alpha G^{5\alpha} \\
&= \tfrac{1}{k} \partial_\alpha G^{5\alpha} + \tfrac{1}{k} \Gamma_\alpha{}^5{}_\beta G^{\beta\alpha} + \tfrac{1}{k} \Gamma_\alpha{}^\alpha{}_\beta G^{5\beta} \\
&= \tfrac{1}{k} \partial_a G^{5a} + \tfrac{1}{k} \partial_5 G^{55} + \tfrac{1}{k} \Gamma_a{}^5{}_c G^{ca} + \tfrac{1}{k} \Gamma_a{}^5{}_5 G^{5a} + \tfrac{1}{k} \Gamma_5{}^5{}_a G^{a5} + \\
&\quad + \tfrac{1}{k} \Gamma_5{}^5{}_5 G^{55} + \tfrac{1}{k} \Gamma_a{}^a{}_c G^{5c} + \tfrac{1}{k} \Gamma_a{}^a{}_5 G^{55} + \tfrac{1}{k} \Gamma_5{}^5{}_a G^{5a} + \\
&\quad + \tfrac{1}{k} \Gamma_5{}^5{}_5 G^{55} \\
&= -\tfrac{1}{k} S_{\alpha\gamma}{}^\beta R^5{}_\beta{}^{\gamma\alpha} - \tfrac{2}{k} S_\alpha{}^{5\beta} R_\beta{}^\alpha \\
&= -\tfrac{1}{k} S_{ab}{}^c R^5{}_c{}^{ba} - \tfrac{1}{k} S_{ab}{}^5 R^5{}_5{}^{ba} - \tfrac{1}{k} S_{a5}{}^b R^5{}_b{}^{5a} - \tfrac{1}{k} S_{a5}{}^5 R^5{}_5{}^{5a} - \\
&\quad - \tfrac{1}{k} S_{5a}{}^b R^5{}_b{}^{a5} - \tfrac{1}{k} S_{5a}{}^5 R^5{}_5{}^{a5} - \tfrac{1}{k} S_{55}{}^a R^5{}_a{}^{55} - \\
&\quad - \tfrac{1}{k} S_{55}{}^5 R^5{}_5{}^{55} - \tfrac{2}{k} S_a{}^{5b} R_b{}^a - \tfrac{2}{k} S_a{}^{55} R_5{}^a - \tfrac{2}{k} S_5{}^{5a} R_a{}^5 - \\
&\quad - \tfrac{2}{k} S_5{}^{55} R_5{}^5 \\
&= -\tfrac{1}{k} S_{ab}{}^c R^5{}_c{}^{ba} - \tfrac{1}{k} S_{ab}{}^5 R^5{}_5{}^{ba} - \tfrac{1}{k} S_{a5}{}^b R^5{}_b{}^{5a} - \tfrac{1}{k} S_{a5}{}^5 R^5{}_5{}^{5a} - \\
&\quad - \tfrac{1}{k} S_{5a}{}^b R^5{}_b{}^{a5} - \tfrac{1}{k} S_{5a}{}^5 R^5{}_5{}^{a5} - \tfrac{2}{k} S_a{}^{5b} R_b{}^a - \tfrac{2}{k} S_a{}^{55} R_5{}^a - \\
&\quad - \tfrac{2}{k} S_5{}^{5a} R_a{}^5 - \tfrac{2}{k} S_5{}^{55} R_5{}^5,
\end{aligned}
$$

respectively.

### 5-Dimensional Tidal Force Density Vector

The two parts of the 5-dimensional tidal force density vector as defined (using anholonomic coordinates for a general linear connection) by

$$
\begin{aligned}
F_{(2)}{}^\alpha &= \mu\, u^\beta \nabla_\beta u^\gamma \nabla_\gamma v^\alpha \qquad (111) \\
&= \mu\, u^\beta v^\gamma u^\delta (R_{\beta\gamma\delta}{}^\alpha - 2\, \nabla_\beta S_{\gamma\delta}{}^\alpha) + 2\, \mu\, u^\beta u^\gamma S_{\beta\delta}{}^\alpha \nabla_\gamma v^\delta \\
&= \mu\, u^\beta v^\gamma u^\delta (R_{\beta\gamma\delta}{}^\alpha - 2\, \nabla_\beta S_{\gamma\delta}{}^\alpha - 4\, S_{\beta e}{}^\alpha S_{\gamma\delta}{}^e) + \\
&\quad + 2\, \mu\, u^\beta v^\gamma S_{\beta\delta}{}^\alpha \nabla_\gamma u^\delta
\end{aligned}
$$

are given, using Eq. (111), by

$$
\begin{aligned}
F_{(2)}{}^a &= \qquad (112) \\
&= \mu\, u^\alpha \nabla_\alpha u^\beta \nabla_\beta v^a \\
&= \mu\, u^b \nabla_b u^c \nabla_c v^a + \mu\, u^b \nabla_b u^5 \nabla_5 v^a + \mu\, u^5 \nabla_5 u^b \nabla_b v^a + \\
&\quad + \mu\, u^5 \nabla_5 u^5 \nabla_5 v^a \\
&= \mu\, u^\alpha v^\beta u^\gamma (R_{\alpha\beta\gamma}{}^a - 2\, \nabla_\alpha S_{\beta\gamma}{}^a) + 2\, \mu\, u^\alpha u^\beta S_{\alpha\gamma}{}^a \nabla_\beta v^\gamma \\
&= \mu\, u^b v^c u^d (R_{bcd}{}^a - 2\, \nabla_b S_{cd}{}^a) + \mu\, u^b v^c u^5 (R_{bc5}{}^a - 2\, \nabla_b S_{c5}{}^a) + \\
&\quad + \mu\, u^b v^5 u^c (R_{b5c}{}^a - 2\, \nabla_b S_{5c}{}^a) + \mu\, u^b v^5 u^5 R_{b55}{}^a + \\
&\quad + \mu\, u^5 v^b u^c (R_{5bc}{}^a - 2\, \nabla_5 S_{bc}{}^a) + \mu\, u^5 v^b u^5 (R_{5b5}{}^a - 2\, \nabla_5 S_{b5}{}^a) - \\
&\quad - 2\, \mu\, u^5 v^5 u^b \nabla_5 S_{5b}{}^a + 2\, \mu\, u^b u^c S_{bd}{}^a \nabla_c v^d + 2\, \mu\, u^b u^c S_{b5}{}^a \nabla_c v^5 + \\
&\quad + 2\, \mu\, u^b u^5 S_{bc}{}^a \nabla_5 v^c + 2\, \mu\, u^b u^5 S_{b5}{}^a \nabla_5 v^5 + \\
&\quad + 2\, \mu\, u^5 u^b S_{5c}{}^a \nabla_b v^c + 2\, \mu\, u^5 u^5 S_{5b}{}^a \nabla_5 v^b \\
&= \mu\, u^\alpha v^\beta u^\gamma (R_{\alpha\beta\gamma}{}^a - 2\, \nabla_\alpha S_{\beta\gamma}{}^a - 4\, \mu\, S_{\alpha\delta}{}^a S_{\beta\gamma}{}^\delta) + 2\, \mu\, u^\alpha v^\beta S_{\alpha\gamma}{}^a \nabla_\beta u^\gamma \\
&= \mu\, u^b v^c u^d (R_{bcd}{}^a - 2\, \nabla_b S_{cd}{}^a) + \mu\, u^b v^c u^5 (R_{bc5}{}^a - 2\, \nabla_b S_{c5}{}^a) + \\
&\quad + \mu\, u^b v^5 u^c (R_{b5c}{}^a - 2\, \nabla_b S_{5c}{}^a) + \mu\, u^b v^5 u^5 R_{b55}{}^a + \\
&\quad + \mu\, u^5 v^b u^c (R_{5bc}{}^a - 2\, \nabla_5 S_{bc}{}^a) + \mu\, u^5 v^b u^5 (R_{5b5}{}^a - 2\, \nabla_5 S_{b5}{}^a) - \\
&\quad - 2\, \mu\, u^5 v^5 u^b \nabla_5 S_{5b}{}^a - 4\, \mu\, u^b v^c u^d S_{be}{}^a S_{cd}{}^e - \\
&\quad - 4\, \mu\, u^b v^c u^d S_{b5}{}^a S_{cd}{}^5 - 4\, \mu\, u^b v^c u^5 S_{bd}{}^a S_{c5}{}^d - \\
&\quad - 4\, \mu\, u^b v^c u^5 S_{b5}{}^a S_{c5}{}^5 - 4\, \mu\, u^b v^5 u^c S_{bd}{}^a S_{5c}{}^d - \\
&\quad - 4\, \mu\, u^b v^5 u^c S_{b5}{}^a S_{5c}{}^5 - 4\, \mu\, u^b v^5 u^5 S_{bc}{}^a S_{55}{}^c - \\
&\quad - 4\, \mu\, u^b v^5 u^5 S_{b5}{}^a S_{55}{}^5 - 4\, \mu\, u^5 v^b u^c S_{5d}{}^a S_{bc}{}^d -
\end{aligned}
$$

$$- 4\,\mu\,u^5\,v^b\,u^5\,S_{5c}{}^a\,S_{b5}{}^c - 4\,\mu\,u^5\,v^5\,u^b\,S_{5c}{}^a\,S_{5b}{}^c +$$
$$+ 2\,\mu\,u^b\,v^c\,S_{bd}{}^a\,\nabla_c\,u^d + 2\,\mu\,u^b\,v^c\,S_{b5}{}^a\,\nabla_c\,u^5 + 2\,\mu\,u^b\,v^5\,S_{bc}{}^a\,\nabla_5\,u^c +$$
$$+ 2\,\mu\,u^b\,v^5\,S_{b5}{}^a\,\nabla_5\,u^5 + 2\,\mu\,u^5\,v^b\,S_{5c}{}^a\,\nabla_b\,u^c + 2\,\mu\,u^5\,v^5\,S_{5b}{}^a\,\nabla_5\,u^b$$

and

$$F_{(2)}{}^5 = \tag{113}$$
$$= \mu\,u^\alpha\,\nabla_\alpha\,u^\beta\,\nabla_\beta\,v^5$$
$$= \mu\,u^a\,\nabla_a\,u^b\,\nabla_b\,v^5 + \mu\,u^a\,\nabla_a\,u^5\,\nabla_5\,v^5 + \mu\,u^5\,\nabla_5\,u^a\,\nabla_a\,v^5 +$$
$$+ \mu\,u^5\,\nabla_5\,u^5\,\nabla_5\,v^5$$
$$= \mu\,u^\alpha\,v^\beta\,u^\gamma\,(R_{\alpha\beta\gamma}{}^5 - 2\,\nabla_\alpha\,S_{\beta\gamma}{}^5) + 2\,\mu\,u^\alpha\,u^\beta\,S_{\alpha\gamma}{}^5\,\nabla_\beta\,v^\gamma$$
$$= \mu\,u^a\,v^b\,u^c\,(R_{abc}{}^5 - 2\,\nabla_a\,S_{bc}{}^5) + \mu\,u^a\,v^b\,u^5\,(R_{ab5}{}^5 - 2\,\nabla_a\,S_{b5}{}^5) +$$
$$+ \mu\,u^a\,v^5\,u^b\,(R_{a5b}{}^5 - 2\,\nabla_a\,S_{5b}{}^5) + \mu\,u^a\,v^5\,u^5\,R_{a55}{}^5 +$$
$$+ \mu\,u^5\,v^a\,u^b\,(R_{5ab}{}^5 - 2\,\nabla_5\,S_{ab}{}^5) + \mu\,u^5\,v^a\,u^5\,(R_{5a5}{}^5 - 2\,\nabla_5\,S_{a5}{}^5) -$$
$$- 2\,\mu\,u^5\,v^5\,u^a\,\nabla_5\,S_{5a}{}^5 + 2\,\mu\,u^a\,u^b\,S_{ac}{}^5\,\nabla_b\,v^c +$$
$$+ 2\,\mu\,u^a\,u^b\,S_{a5}{}^5\,\nabla_b\,v^5 + 2\,\mu\,u^a\,u^5\,S_{ab}{}^5\,\nabla_5\,v^b +$$
$$+ 2\,\mu\,u^a\,u^5\,S_{a5}{}^5\,\nabla_5\,v^5 + 2\,\mu\,u^5\,u^a\,S_{5b}{}^5\,\nabla_a\,v^b + 2\,\mu\,u^5\,u^5\,S_{5a}{}^5\,\nabla_5\,v^a$$

$$= \mu\,u^\alpha\,v^\beta\,u^\gamma\,(R_{\alpha\beta\gamma}{}^5 - 2\,\nabla_\alpha\,S_{\beta\gamma}{}^5 - 4\,S_{\alpha\delta}{}^5\,S_{\beta\gamma}{}^\delta) + 2\,\mu\,u^\alpha\,v^\beta\,S_{\alpha\gamma}{}^5\,\nabla_\beta\,u^\gamma$$
$$= \mu\,u^a\,v^b\,u^c\,(R_{abc}{}^5 - 2\,\nabla_a\,S_{bc}{}^5) + \mu\,u^a\,v^b\,u^5\,(R_{ab5}{}^5 - 2\,\nabla_a\,S_{b5}{}^5) +$$
$$+ \mu\,u^a\,v^5\,u^b\,(R_{a5b}{}^5 - 2\,\nabla_a\,S_{5b}{}^5) + \mu\,u^a\,v^5\,u^5\,R_{a55}{}^5 +$$
$$+ \mu\,u^5\,v^a\,u^b\,(R_{5ab}{}^5 - 2\,\nabla_5\,S_{ab}{}^5) + \mu\,u^5\,v^a\,u^5\,(R_{5a5}{}^5 - 2\,\nabla_5\,S_{a5}{}^5) -$$
$$- 2\,\mu\,u^5\,v^5\,u^a\,\nabla_5\,S_{5a}{}^5 - 4\,\mu\,u^a\,v^b\,u^c\,S_{ae}{}^5\,S_{bc}{}^e -$$
$$- 4\,\mu\,u^a\,v^b\,u^c\,S_{a5}{}^5\,S_{bc}{}^5 - 4\,\mu\,u^a\,v^b\,u^5\,S_{ac}{}^5\,S_{b5}{}^c -$$
$$- 4\,\mu\,u^a\,v^b\,u^5\,S_{a5}{}^5\,S_{b5}{}^5 - 4\,\mu\,u^a\,v^5\,u^b\,S_{ac}{}^5\,S_{5b}{}^c -$$
$$- 4\,\mu\,u^a\,v^5\,u^b\,S_{a5}{}^5\,S_{5b}{}^5 - 4\,\mu\,u^a\,v^5\,u^5\,S_{ab}{}^5\,S_{55}{}^b -$$
$$- 4\,\mu\,u^a\,v^5\,u^5\,S_{a5}{}^5\,S_{55}{}^5 - 4\,\mu\,u^5\,v^a\,u^b\,S_{5c}{}^5\,S_{ab}{}^c -$$
$$- 4\,\mu\,u^5\,v^a\,u^5\,S_{5b}{}^5\,S_{a5}{}^b - 4\,\mu\,u^5\,v^5\,u^a\,S_{5b}{}^5\,S_{5a}{}^b +$$
$$+ 2\,\mu\,u^a\,v^b\,S_{ac}{}^5\,\nabla_b\,u^c + 2\,\mu\,u^a\,v^b\,S_{a5}{}^5\,\nabla_b\,u^5 +$$
$$+ 2\,\mu\,u^a\,v^5\,S_{ab}{}^5\,\nabla_5\,u^b + 2\,\mu\,u^a\,v^5\,S_{a5}{}^5\,\nabla_5\,u^5 +$$
$$+ 2\,\mu\,u^5\,v^a\,S_{5b}{}^5\,\nabla_a\,u^b + 2\,\mu\,u^5\,v^5\,S_{5a}{}^5\,\nabla_5\,u^a,$$

respectively.